\documentclass[numberedappendix,preprint2,longnamesfirst,iop]{emulateapj}
\usepackage{amsmath,graphicx,color,verbatim}

\newcommand{\comments}[1]{}

\newcommand{\unit}[1]{\ensuremath{\,\, \mathrm{#1}}}

\newcommand{\vvir}{\ensuremath{V_{\mathrm{vir}}}}

\newcommand{\svir}{\ensuremath{S_{\mathrm{vir}}}}
\newcommand{\rcl}{\ensuremath{R_{\mathrm{cl}}}}
\newcommand{\rdotcl}{\ensuremath{\dot{R}_{\rm cl}}}

\newcommand{\vcl}{\ensuremath{V_{\mathrm{cl}}}}
\newcommand{\mcl}{\ensuremath{M_{\mathrm{cl}}}}

\newcommand{\mdotacc}{\ensuremath{\dot{M}_{\mathrm{acc}}}}
\newcommand{\mdotej}{\ensuremath{\dot{M}_{\mathrm{ej}}}}
\newcommand{\sigmacl}{\ensuremath{\sigma_{\mathrm{cl}}}}
\newcommand{\vturb}{\ensuremath{\mathbf{v}_{\rm turb}}}
\newcommand{\rhodot}{\ensuremath{\dot{\rho}}}
\newcommand{\rhodotacc}{\ensuremath{\dot{\rho}_{\rm acc}}}
\newcommand{\rhodotej}{\ensuremath{\dot{\rho}_{\rm ej}}}
\newcommand{\aone}{\ensuremath{a_{\textsc{I}}}}

\newcommand{\krho}{\ensuremath{k_\rho}}

\newcommand{\mres}{\ensuremath{M_{\rm res}}}

\newcommand{\vres}{\ensuremath{\mathbf{v}_{\rm res}}}
\newcommand{\vressys}{\ensuremath{v_{\rm res,sys}}}
\newcommand{\vresturb}{\ensuremath{\mathbf{v}_{\rm res,turb}}}
\newcommand{\sigmares}{\ensuremath{\sigma_{\rm res}}}

\newcommand{\vw}{\ensuremath{\mathbf{v}_{\rm w}}}
\newcommand{\vej}{\ensuremath{\mathbf{v}_{\rm ej}^\prime}}
\newcommand{\cii}{\ensuremath{c_\textsc{ii}}}

\newcommand{\fstar}{\ensuremath{\mathbf{F}_*}}

\newcommand{\gv}[1]{\ensuremath{\mbox{\boldmath$ #1 $}}} 
\renewcommand{\div}[1]{\gv{\nabla} \cdot #1} 
\newcommand{\grad}[1]{\gv{\nabla} #1} 
\newcommand{\abs}[1]{\left| #1 \right|} 
\newcommand{\avg}[1]{\left< #1 \right>} 

\newcommand{\hii}{H~\textsc{ii}}
\newcommand{\hi}{H~\textsc{i}}

\shorttitle{Global Evolution of GMCs: Accretion}
\shortauthors{Goldbaum et al.}

\begin{document}

\title{The Global Evolution of Giant Molecular Clouds II:\\ The Role of Accretion}

\author{Nathan J. Goldbaum\altaffilmark{1}, Mark R. Krumholz\altaffilmark{1}, Christopher D. Matzner\altaffilmark{2}, Christopher F. McKee\altaffilmark{3}}

\altaffiltext{1}{Department of Astronomy, 201 Interdisciplinary Sciences Building, University of California, Santa Cruz, CA, 95064, USA; goldbaum@ucolick.org}
\altaffiltext{2}{Department of Astronomy and Astrophysics, University of Toronto, Toronto, ON M5S 3H8, Canada}
\altaffiltext{3}{Physics Department and Astronomy Department, University of California at Berkeley, Berkeley, CA 94720}
\begin{abstract}
We present virial models for the global evolution of giant molecular clouds.  Focusing on the presence of an accretion flow, and accounting for the amount of mass, momentum, and energy supplied by accretion and star formation feedback, we are able to follow the growth, evolution, and dispersal of individual giant molecular clouds.  Our model clouds reproduce the scaling relations observed in both galactic and extragalactic clouds.  We find that accretion and star formation contribute roughly equal amounts of turbulent kinetic energy over the lifetime of the cloud.  Clouds attain virial equilibrium and grow in such a way as to maintain roughly constant surface densities, with typical surface densities of order $50 \textrm{ -- } 200 \unit{M}_\odot \textrm{ pc}^{-2}$, in good agreement with observations of giant molecular clouds in the Milky Way and nearby external galaxies.  We find that as clouds grow, their velocity dispersion and radius must also increase, implying that the linewidth-size relation constitutes an age sequence.  Lastly, we compare our models to observations of giant molecular clouds and associated young star clusters in the LMC and find good agreement between our model clouds and the observed relationship between H~\textsc{ii} regions, young star clusters, and giant molecular clouds.
\end{abstract}
\keywords{stars: formation --- ISM: clouds --- ISM: evolution --- turbulence}

\section{Introduction}

Giant molecular clouds (GMCs) are the primary reservoir of molecular gas in the galaxy \citep{wm97,ros05,sl06}.  Since the surface density of star formation shows a strong correlation with the surface density of molecular gas \citep{wb02,ken07,big08,sch11}, GMCs must also be the primary site of star formation in the Milky Way.  However, recent high-resolution observations have shown that the Kennicutt-Schmidt law breaks down when the resolution of an observation is finer than the typical length scales of GMCs \citep{ono10,sch10}.  Thus, in order to develop a detailed theoretical understanding of the relationship between star formation and molecular gas, it is necessary to first understand the formation, evolution, and destruction of giant molecular clouds.  

One stumbling block in this effort is the substantial disagreement in the literature regarding both the formation mechanism and typical lifetimes of GMCs \citep[see e.g.][and references therein]{gk74,ze74,bs80,bp99,mck07,mur10}.  Some authors suggest that GMCs form out of bound atomic gas as a result of gravitational instability \citep{kim02,kim03,ko06,li06,tt09}, surviving as roughly virialized objects  for many cloud dynamical times \citep{tan06,tam08}.  In support of this picture is the observation that massive clouds are found to be marginally bound, with typical virial parameters of order unity \citep{hey01,ros07,rom10}.  Since supersonic isothermal turbulence is found to decay via radiative shocks in one or two crossing times \citep{mac98,sto98,mlk04,es04}, this model must invoke a mechanism to drive supersonic motions for the lifetime of a cloud, which could be several crossing times.  Possible turbulent driving mechanisms include protostellar outflows \citep{ns80,mck89,lin06,li10}, H~\textsc{ii} regions \citep[hereafter KM09]{mat02,km09}\defcitealias{km09}{KM09}, supernovae \citep{mlk04}, or, as investigated here, mass accretion \citep{kh10,vaz10}.  

Accretion driven turbulence in molecular clouds has received little attention in the literature.  However, as \citet{kh10} point out, the kinetic energy of accreted material can power the turbulent motions observed in molecular clouds with energy conversion efficiencies of only a few percent.  While there has been no systematic study of the kinetic energy budget of a molecular cloud formed via gravitational instability, this problem has been examined in the context of the formation of protogalaxies at high redshift.  In one example, \citet{wa07} analyzed simulations of virializing high redshift minihalos, tracking the thermal, kinetic, and gravitational potential energy of gas in protogalactic dark matter halos.  In their models, which included a nonequilibrium cooling model, gas collapsed onto the halo and cooled quickly, causing turbulent velocities to become supersonic.  As pointed out by \citet{wa08}, this means that the virialization process is a local one: gravitational potential energy can be converted directly into supersonically turbulent motions characterized by a volume-filling network of shocks.  The turbulence in turn provides much of the kinetic support for the newly virialized gaseous component of the dark matter halo.  If a similar mechanism is at work as gas cools and collapses onto GMCs then gravitational potential energy alone may be sufficient to power turbulence in GMC.

The most detailed simulations of simultaneously accreting and star forming giant molecular clouds were recently completed by \citet{vaz10}.  These numerical models included a simplified subgrid prescription for stellar feedback by the ionizing radiation of newborn star clusters and focused on the balance between accretion and feedback in clouds formed via thermal instability in colliding flows.  Throughout the course of the simulation, dense molecular gas condensed out of a warm atomic envelope, allowing a study of the interplay between accretion and feedback in the simulated clouds.  The resulting clouds were able to attain a state of quasi-virial equilibrium, in which the supply of gas from the ambient medium balanced the formation of stars and ejection of gas from {\hii} regions.  Due to the idealized nature of the subgrid star formation feedback prescription, in which all {\hii} regions were powered by a cluster with the same ionizing luminosity, star formation feedback was unable to act on the cloud as a whole but could reduce the global star formation rate by destroying overdensities.  Since the simulation did not include star clusters with large ionizing luminosities, the cloud as a whole could not be destroyed and star formation would have eventually consumed all of the gas had the simulation not been cut off.  Even though the simulations employed a highly idealized star formation prescription, the computations still required substantial resources to complete and only allowed insight into the evolution of a single cloud.  It seems that a computationally inexpensive model that includes a somewhat more sophisticated treatment of star formation feedback is called for.

In this work, we model the global evolution of giant molecular clouds from their birth as low-mass seed clouds to their dispersal after a phase of massive star formation.  This is done using an updated version of the semianalytical model of \citet[][hereafter Paper I]{kmm06}.  Using a virial formalism, we compute the global dynamical evolution of a single cloud while simultaneously tracking its energy budget.  Model clouds form stars, launch H~\textsc{ii} regions and undergo accretion from their environments.   With these models, we are able to investigate the role accretion plays in maintaining turbulence in molecular clouds and directly compare to observations of GMCs in the Milky Way and nearby external galaxies.  This work is complementary to the simulations of \citet{vaz10}, since our simplified global models allow us to survey a large variety of GMCs at little computational cost while including a much more sophisticated star formation feedback prescription.  We are able to capture model clouds with masses comparable to the most massive clouds observed in the Milky Way and nearby galaxies, allowing us to simulate the sites of the majority of star formation in these systems \citep{wm97,fk10}.  

We proceed by describing the formulation and implementation of our GMC model in \S\,\ref{gmcevolution}. Next, in \S\,\ref{varphivary}, we test our implementation of accretion.  Following this, in \S\,\ref{fullsims} we perform full simulations and describe the gneral features of our simulated clouds.  In \S\,\ref{obscompare} we make comparisons to observations, focusing on the scaling relations observed to hold for GMCs as well as the high quality multiwavelength observations available for GMCs in the LMC.  Lastly, in \S\,\ref{caveats}, we discuss the limitations inherent in the simplifying assumptions we make to derive the cloud evolution equations.

\section{Governing Equations}
\label{gmcevolution}

The GMC evolution model described below allows us to solve for the time-evolution of the global properties of model molecular clouds.  In contrast with previous work, we follow the flow of gas as it condenses out of the diffuse gas in the envelope surrounding the GMC and falls onto the cloud.  Employing simplifying assumptions as well as the results of simulations of compressible MHD turbulence, we derive a set of coupled ordinary differential equations that govern the time evolution of the cloud's mass, radius, and velocity dispersion.  Combining the governing evolution equations with a set of initial conditions, model parameters, and a model for the time dependence of the mass accretion rate based on the gravitational collapse of the GMC envelope, we can solve for the time evolution of the cloud.  Below, we give an overview of the model, discuss the formulation of our numerical scheme, describe our parameter choices, and give a brief description of our treatment of star formation and our model for the GMCs gas supply.

\subsection{Model Overview}
\label{overview}

\begin{figure}
\resizebox{\columnwidth}{!}{\includegraphics{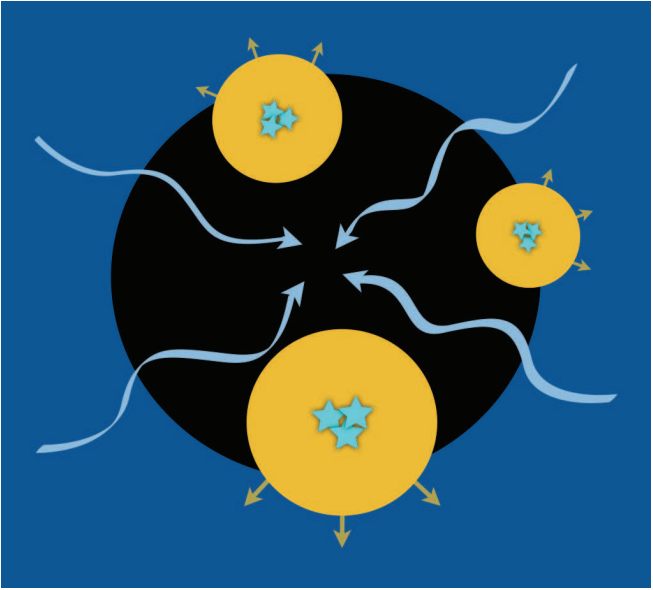}}
\caption{A schematic overview of the GMC model.  A molecular cloud ({\it black}) is embedded in a warm atomic envelope ({\it dark blue}).  Cool atomic gas ({\it light blue}) flows onto the cloud, where it condenses, recombines into molecules, and mixes with the cloud.  Newborn OB associations ({\it blue stars}) drive {\hii} regions ({\it orange}) and eject winds ionized winds back into the ambient medium.}
\label{modelcartoon} 
\end{figure}

The model we employ below is based on the global GMC model of Paper I, itself a generalization of the  the global model for low mass star formation of \citet{mck89}, the Eulerian virial theorem (EVT) of \citet{mz92}, and the model for star forming clumps of \citet{mat01}.  Employing a virial formalism, we account for the dynamics and energy budget of gas contained within an Eulerian volume, $V_{\rm vir}$.  We separate the gas within $\vvir$ into three species: virial material, a gaseous reservoir, and a photoionized wind.  A schematic representation of the components of our model is presented in Figure~\ref{modelcartoon}.

By design, each of the three components has a straightforward physical interpretation.  The first component, which we label virial material, consists of two physically distinct subcomponents: a molecular cloud and a warm atomic envelope that encloses the cloud.  The cloud is assumed to be cold ($\sim 10\unit{K}$), molecular, and contained within a spherical volume of radius $\rcl$.  The ambient medium is composed of warm ($\sim 10^3 \unit{K}$) and diffuse atomic gas that encloses the cloud and extends beyond the virial volume.  The second component is a gaseous reservoir, which we assume is composed of cold ($\sim 10^2 \unit{K}$) neutral material that flows onto the cloud at free-fall from beyond the virial volume.  The last component is an ionized wind made up of hot ($\sim 10^4 \unit{K}$) ionized gas ejected from the ionization fronts of blister-type {\hii} regions.  All three components are allowed to mutually interpenetrate.  We restrict interaction between the components to the transfer of mass between the accretion flow and cloud as well as between the cloud and wind.  Since the envelope and cloud are not allowed to interpenetrate,  we formally group the envelope and cloud together; this somewhat artificial choice significantly simplifies the virial analysis.

We make use of two simplifying assumptions regarding the distribution and flow of the virial material.  First, we assume that the virial material follows a spherically symmetric, smoothly varying density profile.  Second, we assume that the the cloud is homologous: the cloud expands, contracts, accretes, and sheds mass in such a way as to always maintain the same smooth density profile.  We assume a density profile of the form
\begin{equation}
\rho(r) = \rho_0\left(\frac{r}{R_{\rm cl}}\right)^{-k_\rho} \text{ for } r\le \rcl
\label{densprofile}
\end{equation}

\noindent
where $\rho_0$ is the density at the edge of the cloud and $k_\rho$ is assumed to be unity.  This choice is consistent with the Larson scaling relations observed in galactic \citep{lar81,sol87,hey01,hey09} and extragalactic \citep{miz01,eng03,ros07,bol08,hug10} GMCs. Our assumed cloud density profile effectively averages over clumpy and filamentary internal structure and oblate shapes of observed clouds.  At $r = R_{\rm cl}$, the density of the ambient medium is assumed to smoothly transition from $\rho_0$ to $\rho_{\rm amb}$ in a thin boundary layer.  We assume that the density of the ambient medium is negligible compared to the density of gas in the cloud, $\rho_{\rm amb} \ll \rho_0$.  The density of the ambient medium remains $\rho_{\rm amb}$ out to the virial radius.

Beyond assuming a density profile, we must also specify the velocity structure of all three components of the model.  We follow Paper I in assuming that the velocity of the virial material can be decomposed into a systematic and turbulent component,
\begin{equation}
\mathbf{v} = \frac{\rdotcl}{\rcl}\mathbf{r} + \vturb.
\end{equation}

\noindent
We assume that $\vturb$ is randomly oriented with respect to position so that turbulent motions carry no net flux of matter.  We make a similar assumption regarding the velocity structure of the reservoir,
\begin{equation}
\vres = \vressys \hat{r} + \vresturb.
\end{equation}

\noindent
The systematic component of $\vres$ is due to the gravitational attraction of material within the virial volume
\begin{equation}
\frac{\vressys^2}{2} = \int_\infty^r \mathbf{g} \cdot d\mathbf{r},
\label{vressysdef}
\end{equation}

\noindent
while the random component is such that $(\mcl^{-1}\int_{\vcl} \rho \mathbf{v^\mathrm{2}_{\rm res,turb}} dV)^{1/2} = \sqrt{3}\sigma_{\rm res}$.  Here $\mathbf{g}$ is the gravitational acceleration and $\sigmares$ is the velocity dispersion of gas in the reservoir feeding the accretion flow.  Since the amount of material in the accretion flow is determined by how fast it can fall into the cloud, we must simultaneously determine both the density profile and radial velocity of material in the accretion flow (see Appendix~\ref{accretionappendix}).  Finally, for the wind material, we follow Paper I in assuming

\begin{equation}
\vw = \mathbf{v} + \vej
\end{equation}

\noindent
where $\vej = 2\cii\hat{r}$ and $\cii$ is the ionized gas sound speed.  We follow \citet{mw97} in choosing $\cii = 9.7 \unit{km}\unit{s}^{-1}$.

\subsection{Momentum Equation}
\label{momeqsection}

In Appendix~\ref{derivationEVT}, we derive the EVT for a simultaneously evaporating and accreting cloud,
\begin{equation}\begin{split}
\frac{1}{2}\ddot{I}_{\rm cl} &= 2(\mathcal{T} - \mathcal{T}_0) + \mathcal{B} + \mathcal{W} - \frac{1}{2}\frac{d}{dt}\int_{S_{\rm vir}} (\rho \mathbf{v} r^2) \cdot d\mathbf{S} \\
&+ a_\textrm{I}\dot{M}_{\rm cl}R_{\rm cl}\dot{R}_{\rm cl} + \frac{1}{2}a_\textrm{I}\ddot{M}_{\rm cl}R_{\rm cl}^2 + a_\textrm{I}\dot{M}_{\rm ej}R_{\rm cl}\dot{R}_{\rm cl} \\
&+ \frac{3-k_\rho}{4-k_\rho}R_{\rm cl}(\dot{M}_{\rm ej}v_{\rm ej}^\prime - \xi\dot{M}_{\rm acc}v_{\rm esc})
\end{split}\end{equation} 
Here, {\aone} is a constant of order unity that depends on the distribution of material in the cloud, $I_{\rm cl}$ is the cloud moment of inertia, $\mathcal{T}$ is the combined turbulent and thermal kinetic energy of the cloud, $\mathcal{T}_0$ is the energy associated with interstellar pressure at the cloud surface, $\mathcal{B}$ is the net magnetic energy due to the presence of the cloud, $\mathcal{W}$ is the gravitational term (equal to the gravitational binding energy in the absence of an external potential [\citealt{mz92}]), the surface integral is proportional to the rate of change of the moment of inertia inside the bounding viral surface, $M_{\rm cl}$ is the cloud mass, $R_{\rm cl}$ is the cloud radius, $\dot{M}_{\rm ej}$ is the mass ejection rate, $\mdotacc$ is the mass accretion rate, $v_{\rm esc} = \{2G [M_{\rm cl}+M_{\rm res}(R_{\rm cl})]/ R_{\rm cl}\}^{1/2}$ is the escape velocity at the edge of the cloud, and $\xi$ is a dimensionless factor we compute via equation~(\ref{xidef}) that depends on the depth of the cloud potential well.  The quantity $\xi v_{\rm esc}$ is the accretion rate weighted average infall velocity.  Precise definitions for $\mathcal{T}$, $\mathcal{T}_0$, $\mathcal{B}$, and $\mathcal{W}$ are given in Paper I.  

The EVT of a cloud without accretion or mass loss would only contain the terms up to the surface integral.  The next three terms account for changes in the cloud moment of inertia due to changes in the mass of the cloud, while the last term accounts for the rate at which the recoil of inflowing and outflowing mass injects momentum into the cloud.   Inflows and outflows are treated separately because material is ejected at a constant velocity, but is accreted at a velocity that is a function of the depth of the potential well of the cloud.  The dimensionless factor $\xi$ appears due to this difference.

The mass of the cloud can only change via mass accretion or ejection,
\begin{equation}
\dot{M}_{\rm cl} = \dot{M}_{\rm ej} + \dot{M}_{\rm acc}.
\label{mdot_def}
\end{equation}

\noindent
We assume that ejection of material can only decrease the mass of the cloud and accretion can only increase the mass of the cloud.  Since stars may not follow the homologous density profile we assume, we neglect the change in the cloud mass due to star formation.  We expect the error incurred from this assumption will be small, since stars make up a small fraction of the mass of observed clouds \citep{ev09,lla10} and our assumed star formation law converts a small fraction of the cloud's mass into stars per free-fall time.

We follow Paper I in using the assumption of homology and the results of simulations of MHD turbulence to evaluate each term in the EVT in terms of constants and the dynamical variables $R_{\rm cl}$, $M_{\rm cl}$, and $\sigma_{\rm cl}$.  In the end, we obtain a second order nonlinear ordinary differential equation in $\rcl$,
 \begin{equation}\begin{split}
 a_I \ddot{R}_{\rm cl} = &\,3 \frac{c_{\rm cl}^2}{R_{\rm cl}} + 3.9 \frac{\sigma_{\rm cl}^2}{R_{\rm cl}}  - \frac{3}{5}a^\prime(1 - \eta_B^2)\frac{GM_{\rm cl}}{R_{\rm cl}^2}
 \\& 
 - 4\pi P_{\rm amb}\frac{R_{\rm cl}^2}{M_{\rm cl}} - a_I\frac{\dot{M}_{\rm acc}}{M_{\rm cl}}\dot{R}_{\rm cl}
 \\&
 + \left(\frac{3 - k_\rho}{4 - k_\rho}\right)\left(\frac{\dot{M}_{\rm ej}}{M_{\rm cl}}v^\prime_{\rm ej} - \xi \frac{\dot{M}_{\rm acc}}{M_{\rm cl}}v_{\rm esc}\right).
 \label{reqn1}
\end{split} \end{equation}
Here,
\begin{equation}
a^\prime = \frac{15 - 5\krho}{15-6\krho}\left[1 + (3-2\krho)\int_0^1x^{1-\krho}y(x)dx\right],
\end{equation}
$y(x)$ is the ratio of the mass of reservoir material to the mass of cloud material contained within a normalized radius $x = r/R_{\rm cl}$ (see Appendix~\ref{accretionappendix}), and $\eta_{\rm B}$ is the ratio of the magnetic critical mass to the cloud mass.

This equation governs the balance of forces acting on the cloud as a whole.  Each term corresponds to a single physical mechanism that can alter the radial force balance.  The first two terms are due to thermal and turbulent pressure support, respectively.  The third is due to a combination of gravitational compression and magnetic support.  The fourth is due to the confining interstellar pressure. The fifth comes from the exchange of momentum between the expanding cloud and infalling accretion flow. Finally, the last term is due to a combination of the recoil from ejected material and the ram pressure of accreting material.  Although the two parts of the recoil term have opposite signs, $\dot{M}_{\rm ej}$ and $\dot{M}_{\rm acc}$ have opposite signs as well: $\dot{M}_{\rm ej}<0$ and $\dot{M}_{\rm acc}>0$.  This implies that both the recoil due to launching wind material and the ram pressure of accreting reservoir material tend to confine the cloud.

Letting $M_{\rm cl,0}$, $R_{\rm cl,0}$, and $\sigma_{\rm cl,0}$ be the cloud mass, radius, and velocity dispersion at $t = 0$ and defining the initial cloud crossing time, $t_{\rm cr,0} = R_{\rm cl,0}/\sigma_{\rm cl,0}$, we can define the dimensionless variables $M = M_{\rm cl}/M_{\rm cl,0}$, $R = R_{\rm cl}/R_{\rm cl,0}$, $\sigma = \sigma_{\rm cl}/\sigma_{\rm cl,0}$, and $\tau = t/t_{\rm cr,0}$.  Letting primes denote differentiation with respect to $\tau$, we can write equation~(\ref{reqn1}) in dimensionless form
\begin{equation}\begin{split}
R^{\prime \prime} = &
\frac{3.9\sigma^2 + 3\mathcal{M}_0^{-2}}{a_I R} - \eta_G\frac{a^\prime M}{R^2} - \eta_P\frac{R^2}{M} 
\\&
- \frac{M^\prime_{\rm acc}R^\prime}{M} + \eta_{E}\frac{M^\prime_{\rm ej}}{M} - \eta_{A}\frac{\xi M^\prime_{\rm acc}}{(fMR)^{1/2}}
\label{reqn}
\end{split}
\end{equation}
where 
\begin{equation}
\mathcal{M}_0 = \sigma_{\rm cl,0} / c_{\rm cl}
\end{equation}
is the initial turbulent Mach number and we define the dimensionless constants 
\begin{align}
&\eta_G =\, \frac{3(1 - \eta_B^2)}{a_I\alpha_{\rm vir,0}}\\
&\eta_P =\, \frac{4\pi R_{\rm cl,0}^3P_{\rm amb}}{a_IM_{\rm cl,0}\sigma_{\rm cl,0}^2}\\
&\eta_E =\, \left( \frac{5 - k_\rho}{4 - k_\rho} \right) \frac{v^\prime_{\rm ej}}{ \sigma_{\rm cl,0}}\\
&\eta_A =\, \left( \frac{5 - k_\rho}{4 - k_\rho}\right) \left( \frac{10}{ \alpha_{\rm vir,0}} \right)^{1/2}
\end{align}
where 
\begin{equation}
\alpha_{\rm vir,0} = \frac{5 \sigma_{\rm cl,0}^2R_{\rm cl,0}}{GM_{\rm cl,0}}
\label{virialparam}
\end{equation}
is the initial nonthermal virial parameter \citep{ber92}.  These constants are set by the ratios of various forces acting on the initial state of the cloud.  $\eta_{\rm G}$ is proportional to the ratio of the initial magnetic  forces to the initial gravitational force, and $\eta_P$, $\eta_E$, and $\eta_A$ are the ratios of the ambient pressure force, the mass ejection recoil force, and the initial accretion ram pressure force to the initial internal turbulent forces, respectively.

Comparing equation~(\ref{reqn}) with the corresponding equation given in Paper I, we see that two new terms proportional to $M^\prime_{\rm acc}$ have appeared.  In practice, we find that, of the two terms, the one proportional to $\eta_A$ dominates, implying that the primary direct impact of accretion on the radial force balance of the cloud is to provide a confining ram pressure.  We will see in the next section that accretion also increases the turbulent velocity dispersion, implying that the kinetic pressure term also increases when accretion is included.  The cloud radius is determined by a balance between kinetic pressure and a combination of gravity, accretion ram pressure, and wind recoil pressure.  Thermal pressure support is negligible.

We also note that although the ambient pressure term is of the same form as in Paper I, we assume an observationally motivated value for the ambient pressure, $P_{\rm amb}/k_{\rm B} = 3\times10^4 \unit{K}\unit{cm}^{-3}$ \citep{mck99}.  This includes thermal and turbulent pressure but neglects magnetic and cosmic ray pressure, since magnetic fields and cosmic rays permeate both the cloud and the ambient ISM.  We have also adjusted the ambient pressure upwards by a factor of two because GMCs form in overdense regions of the ISM where the hydrostatic pressure is higher than average.  In Paper I, $P_{\rm amb}$ was chosen to be artificially high to ensure that the cloud would start its evolution in hydrostatic equilibrium.  This choice was made to account for the weight of the gaseous reservoir that was not explicitly included.  In practice, by choosing a lower value for $P_{\rm amb}$, we find that the ambient pressure term is subdominant for most of the evolution of the cloud.  This is expected, since we now correctly account for the pressure of the reservoir through the term proportional to $\eta_{\rm A}$.  Once accretion halts, the cloud is left out of pressure equilibrium and must expand to match the ambient pressure.  This effect is seen most clearly in the second column of Figure~\ref{cloudpropsfeedback}.

\subsection{Energy Equation}
In Appendix~\ref{derivationEnergy}, we derive the time evolution equation for the total energy of the cloud,
\begin{equation}\begin{split}
&\frac{d\mathcal{E}_{\rm cl}}{dt}=\frac{\dot{M}_{\rm cl}}{M_{\rm cl}}[\mathcal{E}_{\rm cl}+(1-\eta_{B}^{2})\mathcal{W}] - 4\pi P_{\rm amb}R_{\rm cl}^{2}\dot{R}_{\rm cl}\\
&+\frac{GM_{\rm cl}\dot{M}_{\rm cl}}{R_{\rm cl}}\chi\left(1 - \frac{M_{\rm cl}\dot{R}_{\rm cl}}{\dot{M}_{\rm cl}R_{\rm cl}}\right) \\
&+ \left(\frac{3-k_{\rho}}{4-k_{\rho}}\right)\dot{R}_{\rm cl}(\dot{M}_{\rm ej}v_{\rm ej}^{'}-\xi\dot{M}_{\rm acc}v_{\rm esc}) \\
&+  \varphi(\frac{3}{2} \dot{M}_{\rm acc}\sigma_{\rm res}^2 + \frac{3}{2}\dot{M}_{\rm acc}\sigma_{\rm cl}^2 + \gamma \dot{M}_{\rm acc} v_{\rm esc}^2) \\
&- a_I\dot{M}_{\rm acc}\dot{R}_{\rm cl}^{2} - 3\dot{M}_{\rm acc}\sigma_{cl}^{2} + \mathcal{G}_{\rm cl}-\mathcal{L}_{\rm cl},
\label{energyeqn1}
\end{split}\end{equation}
where $\mathcal{G}_{\rm cl}$ and $\mathcal{L}_{\rm cl}$ are the rates of energy gain and loss due to H~\textsc{ii} regions and turbulent dissipation, respectively, $\mathcal{E}_{\rm cl}$ is the total energy due to the presence of the cloud (see equation~[\ref{ecldef}]), $\sigma_{\rm res}$ is the velocity dispersion of material that is being accreted, $\chi$ is given by Equation~\ref{chidef}, $\gamma$ is given by equation~(\ref{gammadef}), and $\varphi$ is a free parameter that sets the amount of energy available to drive accretion driven turbulence.  The parameter $\varphi$ is the only adjustable constant in our model that is not constrained by the results of simulations or observations and must be tuned to reproduce the observed properties of clouds.  The evolution of the cloud is very sensitive to $\varphi$ and we justify our fiducial choice, $\varphi = 0.75$, in \S\,\ref{varphivary}.

This equation governs the global energy budget of the cloud.  Each term has a straightforward physical explanation.  The term $(\dot{M}_{\rm cl}/M_{\rm cl})\mathcal{E}_{\rm cl}$ is the rate of change of the cloud's energy as mass is advectively added to or carried away from the cloud.  Similarly, $(\dot{M}_{\rm cl}/M_{\rm cl})(1 - \eta_{\rm B}^2)\mathcal{W}$ is the rate of change of the gravitational and magnetic energy due to changes in the mass of the cloud.  The next term is the rate at which external pressure does compressional work on the cloud.  This is followed by a term that accounts for the gravitational work done on the cloud by the reservoir as the cloud expands and contracts.  The following term represents the rate at which mass inflows, outflows, and external thermal and turbulent pressure, respectively, do compressional work on the cloud.  The next term, which is proportional to $\varphi$, represents the rate of kinetic energy injection via stirring of turbulence by accreted material.  This is followed by two terms that are proportional to $\dot{M}_{\rm acc}$, which account for the fact that in the frame comoving with the motions of material in the cloud, accreted material is moving at the transformed velocity, $\mathbf{v}_{\rm res} - \mathbf{v}$, different than the velocity of the reservoir material in the rest frame, $\mathbf{v}_{\rm res}$.  Lastly, $\mathcal{G}_{\rm cl}$ and $\mathcal{L}_{\rm cl}$ are the rate of energy injection by star formation and the rate at which energy is radiated away, respectively.  
 
Noting that turbulent motions carry no net radial flux of matter and recalling that we had set $\mathcal{B}_{\rm turb} = 0.6\mathcal{T}_{\rm turb}$, we may evaluate equation~(\ref{ecldef}) and obtain for the total cloud energy,
\begin{equation} \begin{split}
\mathcal{E}_{\rm cl} =& \frac{1}{2}a_IM_{\rm cl}\dot{R}_{\rm cl}^2 + 2.4 M_{\rm cl}\sigma_{\rm cl}^2 + \frac{3}{2}M_{\rm cl}c_{\rm cl}^2 \\
&- \left[\frac{3}{5}a^\prime(1 - \eta_B^2) + \chi\right]\frac{GM_{\rm cl}^2}{R_{\rm cl}}.
\label{cloudenergy}
\end{split} \end{equation}
Taking the time derivative of this expression, substituting into equation~(\ref{energyeqn1}), and nondimensionalizing as in \S\,\ref{momeqsection} yields a time evolution equation for $\sigma$,
\begin{equation}\begin{split}
&\frac{4.8}{a_I}\sigma^\prime = -\frac{R^\prime R^{\prime \prime}}{\sigma} - \eta_G \frac{M R^\prime}{R^2 \sigma} - \eta_P \frac{R^2 R^\prime}{M \sigma} 
\\&
+ \eta_{E}\frac{M^\prime_{\rm ej}R^\prime}{M \sigma} - \eta_A\frac{M^\prime_{\rm acc} R^\prime}{(M R)^{1/2}\sigma} - \frac{M^\prime_{\rm acc}R^{\prime2}}{M\sigma} 
\\&
- \frac{(3 - 1.5\varphi)M^\prime_{\rm acc}\sigma}{a_I M} + \eta_D\frac{\varphi \varsigma M^\prime_{\rm acc}}{M \sigma} + \eta_I\frac{\varphi \gamma M^\prime_{\rm acc}}{fR\sigma} \\
&+\frac{\mathcal{G} - \mathcal{L}}{a_I M \sigma} 
\label{sigmaeqn}
\end{split}\end{equation}
where $\varsigma = \sigma_{\rm res}/ \sigma_{\rm res,0}$,
\begin{equation}
\mathcal{G} - \mathcal{L} = \frac{R_{\rm cl,0}(\mathcal{G}_{\rm cl} - \mathcal{L}_{\rm cl})}{M_{\rm cl,0} \sigma_{\rm cl,0}^3},
\end{equation}
and we define the constants,
\begin{align}
\eta_D =& \frac{3\sigma_{\rm res,0}^2}{2a_{\rm I}\sigma_{\rm cl,0}^2}, \\
\eta_I =& \frac{10}{a_{\rm I} \alpha_{\rm vir,0}}.
\end{align}
Here, $\eta_D$ is propotional to the ratio of the initial turbulent kinetic energy in the reservoir and the initial turbulent kinetic energy in the cloud and $\eta_I$ is proportional to the ratio of the initial kinetic energy due to the infall of the reservoir to the initial turbulent kinetic energy of the cloud. 

Since motions in GMCs are highly supersonic, the internal structure of a typical cloud is characterized by strong shocks.  Because clouds have short cooling timescales, the shocks present throughout GMCs must be radiative.  The braking of turbulent motions via radiative shocks has been extensively studied in numerical simulations \citep[see e.g.][]{mac98,sto98} in which the turbulent dissipation timescale is found to be $t_{\rm dis} = E_{\rm turb}/\dot{E} = k \lambda_{\rm in}/\sigma_{\rm cl}$ where $k$ is a constant of order unity and $\lambda_{\rm in}$ is the characteristic length scale of turbulent energy injection.  The simulations of \citet{sto98} give $k = 0.48$ and $E_{\rm turb} = 2.4M_{\rm cl}\sigma_{\rm cl}^{2}$. Motivated by this result and using a scaling argument given by \citet{mat02} and \citet{mck89}, we assume that the dimensionless rate of energy loss is given by
\begin{equation}
\mathcal{L} = \frac{\eta_v}{\phi_{\rm in}}\frac{M \sigma^3}{R}.
\label{turbdecay}
\end{equation}
Here $\eta_v$ is a constant of order unity that depends on the nature of MHD turbulence in the cloud and we assume $\phi_{\rm in} = \lambda_{\rm in} / 4 R_{\rm cl}$.  The factor of 4 in our expression for $\phi_{\rm in}$ comes from the fact that the largest wavelength mode supported by the cloud is $\lambda_{\rm max} = 4R_{\rm cl}$, corresponding to net expansion or compression of the cloud.  We make use of the simulations of \citet{sto98} to calibrate this expression.  For the run most resembling real molecular clouds, we find $\eta_v = 1.2$.  We follow Paper I in adopting $\phi_{\rm in} = 1.0$ below.  This is motivated by the results of  \citet{bru09} \citep[but see also][]{oss02,hb04} who compared the velocity structure of observed clouds, where $\lambda_{\rm in}$ cannot be directly observed, with the velocity structure of simulated clouds, where $\lambda_{\rm in}$ is known {\it a priori}, and found $\lambda_{\rm in} \gtrsim R_{\rm cl}$.

Comparing our velocity dispersion evolution equation (equation~[\ref{sigmaeqn}]) to the corresponding equation given in Paper I, we see there are four new terms proportional to $M^\prime_{\rm acc}$.  In practice, we find that the primary effect of accretion on the energy balance of the cloud is to increase the turbulent velocity dispersion via the terms proportional to $\varphi$.  We will show in \S\,\ref{energetics} that the velocity dispersion is set by a balance between the decay of turbulence and energy injected by accretion and star formation.

\subsection{Star Formation and H~\textsc{ii} Regions}
\label{starformation}

Star formation is able to influence the evolution of the cloud by ejecting mass and by injecting turbulent kinetic energy as expanding {\hii} regions merge with and drive turbulent motions in the cloud.  The first mechanism is accounted for in our models by including an ionized wind that decreases the mass of the cloud and confines the cloud by supplying recoil pressure.    The second mechanism is accounted for by the $\mathcal{G}_{\rm cl}$ term in equation~(\ref{energyeqn1}) that represents energy injection by H~\textsc{ii} regions.

Since we only know the global properties of the cloud, we calculate the rate of star formation by making use of a power-law fit to the star formation law of \citet{km05}.  Stars form at a low efficiency per free-fall time, consistent with observations of star formation in nearby molecular clouds \citep{kt07}.  Individual star formation events occur once a sufficient amount of mass has accumulated to form a star cluster.  Star cluster masses are found by drawing from a cluster mass function appropriate for a single cloud (see equation [44] of Paper I).  We then populate the cluster with individual stars by picking masses from a \citet{kro02} IMF.  If the total ionizing luminosity of the newborn star cluster is sufficient to drive the expansion of an {\hii} region, we begin to track the resulting expansion.

Once a massive star cluster forms, it photoionizes gas in its surroundings and drives the expansion of an {\hii} region.  Paper I tracked the expansion of individual {\hii} regions by assuming the analytic self-similar solution for {\hii} region expansion worked out by \citet{mat02}.  This solution uses the fact that once an {\hii} region has expanded beyond the Str\"{o}mgren radius, most of the mass in the {\hii} region volume is in a thin shell of atomic gas at a radius $r_{\rm sh}$ from the center of the {\hii} region.  The ionized gas in the interior of the shell exerts a pressure on the surface of the shall, causing the shell to accelerate outwards.  The shell evolution equation derived from this analysis admits a self-similar solution for the expansion of the {\hii} region.  This self-similar result does a good job of predicting the expansion if there is no characteristic scale in the problem.  

However, the introduction of radiation pressure leads to a characteristic radius, $r_{\rm ch}$, and time, $t_{\rm ch}$, at which the gas pressure and radiation pressure at the inner surface of the shell are equal.  Radiation pressure is the dominant force driving the expansion of the ionized bubble when $r_{\rm sh} < r_{\rm ch}$ and gas pressure dominates when $r_{\rm sh} > r_{\rm ch}$.  \citetalias{km09} modified the theory of \citet{mat02} to account for the effect of radiation pressure in the initial stages of the expansion.  They derived an explicit functional form for $r_{\rm ch}$ and $t_{\rm ch}$ in terms of the bolometric and ionizing luminosity of the central star cluster, properties of the molecular cloud, and fundamental constants (see equations [4] and [9] in \citetalias{km09}).  The numerical value of $r_{\rm ch}$ and $t_{\rm ch}$ depends on the the bolometric luminosity of the central star cluster and the ionizing photon flux of the central star cluster.  The value we choose for $f_{\rm trap}$, a factor that accounts for the trapping and reradiation of photons as well as the trapping of main sequence winds within the neutral shell, and $\phi$, a factor that accounts for the absorption of radiation by dust, are the fiducial values quoted by \citetalias{km09}.

Defining the dimensionless variables $x_{\rm sh} = r_{\rm sh}/r_{\rm ch}$ and $\tau_{\rm sh} = t / t_{\rm ch}$, the equation of motion for the shell reduces to \citepalias{km09}

\begin{equation}
\frac{d}{d\tau_{\rm sh}}\left( x_{\rm sh}^2 \frac{dx_{\rm sh}}{d\tau_{\rm sh}}\right) = 1 + x_{\rm sh}^{1/2}.
\label{xsheqn}
\end{equation}

\noindent
This assumes that gas in the neighborhood of the {\hii} region follows a density profile proportional to $r^{-1}$, effectively placing the {\hii} region in the center of the cloud.  This accounts for the fact that {\hii} regions form in overdense regions of the cloud.  In practice, we solve equation~(\ref{xsheqn}) numerically to obtain $x_{\rm sh}$ and $d x_{\rm sh}/d\tau_{\rm sh}$ and thus $r_{\rm sh}(t)$ and $\dot{r}_{\rm sh}(t)$.  

In a gas pressure driven {\hii} region, the force exerted on the expanding bubble is twice the recoil force photoionized material imparts on the cloud as it is ejected \citep{mat02}.  The photoevaporation rate can then be straightforwardly calculated via $\mdotej = -\dot{p}_{\rm sh}/{2\cii}$.  Here $p_{\rm sh}$ is the momentum of the shell and $\dot{p}_{\rm sh}$ is the force acting on the shell.  In a radiation pressure driven {\hii} region, the total force is given by the sum of the gas pressure and radiation pressure forces, $\dot{p}_{\rm sh} = \dot{p}_{\rm gas} + \dot{p}_{\rm rad}$.  At early times, when $r_{\rm sh} \ll r_{\rm ch}$, the radiation force dominates, so $\dot{p}_{\rm rad} \gg \dot{p}_{\rm gas}$.  Thus, If we calculate the mass ejection rate from the the total force acting on the shell, we will overestimate the mass ejection rate in a radiation pressure dominated {\hii} region.  To correct for this effect, we modify the analysis of Paper I by only including the gas pressure force when we calculate the mass ejection rate,

\begin{equation}
\mdotej = -\frac{\dot{p}_{\rm gas}}{2\cii}
\end{equation}
where,
\begin{align}
\dot{p}_{\rm gas} =& \left( \frac{12S\phi\pi}{\alpha_{\rm B}} \right)^{1/2} 2.2 k_{\rm B} T_{\textsc{ii}} r_{\rm sh}^{1/2} \\
 =&~ 3.3 \times 10^{28} L_{39}\, x_{\rm sh}^{1/2} \unit{dynes}.
\end{align}

\noindent
Since we do not include the dynamical ejection of material as {\hii} regions break out of the cloud surface, this is formally a lower limit on the true mass ejection rate.  Since clouds are clumpy and somewhat porous \citep{lopez11}, we expect to make little error by neglecting dynamical ejection.

Once the stars providing 50\% of the total ionizing luminosity of the central star cluster have left the main sequence, the {\hii} region enters an undriven momentum-conserving snowplow phase.  When the expansion velocity of the {\hii} region is comparable to the turbulent velocity dispersion, we assume that the {\hii} region breaks up and contributes turbulent kinetic energy to the cloud.  {\hii} regions can merge with the cloud during either the driven or undriven phases.  If the radius of the {\hii} region is greater than the radius of the cloud and the expansion velocity of the shell is greater than the cloud escape velocity, we say the {\hii} region disrupts the cloud and end the global evolution.  

If an {\hii} region merges with the cloud at time $t = t_{m}$ when its radius is $r_{\rm sh} = r_{\rm m}$, the rate of energy injection from a single {\hii} region is given by

\begin{equation}
\mathcal{G}_{\rm cl} = 1.6 \eta_E \mathcal{T}_1(t_m) \left( \frac{r_m}{\rcl} \right)^{1/2} \delta(t - t_m).
\label{energyinjection}
\end{equation}
Here $\mathcal{T}_1 (t_m) = p_{\rm sh} \sigmacl/2$, $\eta_E$ paramaterizes the efficiency of energy injection, $\delta(t)$ is the Dirac delta function, and the factor of 1.6 arises because magnetic turbulence is slightly sub-equipartition compared to kinetic turbulence.  The factor $(r_m/\rcl)^{1/2}$ accounts for the more rapid decay of turbulence when the driving scale is smaller (see Paper I).

\subsection{Mass Accretion}

Consistent with the analysis in Appendix~\ref{accretionappendix}, we treat the reservoir as a gravitationally unstable spherical cloud undergoing collapse.  The cloud is primarily composed of atomic gas in both warm and cold phases.  We expect that as the reservoir collapses, material that is accreted onto the cloud will cool and become molecular.  We approximate that the reservoir has approximately constant surface density, $\Sigma_{\rm res}$.  To find an upper limit on the mass accretion rate, we can assume that the reservoir is undergoing collapse in the limit of zero pressure.  It is straightforward to show that the resulting accretion rate onto the central condensation is given by
\begin{equation}
\label{mdotaccff}
\dot{M}_{\rm acc,ff} = \frac{256}{\pi}G^2 \Sigma_{\rm res}^3 t^3 \\
\end{equation}
Since the estimate for the accretion rate given in equation~\ref{mdotaccff} does not take into account pressure support, it is likely an overestimate of the true accretion rate.  Indeed, \citet{mt03} considered the inside out collapse of equilibrium polytropic molecular cloud cores and, in the case of $\krho = 1$, found the same scaling with time and surface density but a substantially lower coefficient.  \citet{tm04} argued that subsonic inflow was a more realistic initial condition condition than the static equilibrium solution used by \citet{mt03}.  \citet{hun77} found that the set of solutions for the collapse of an isothermal sphere starting at an infinite time in the past.  The solution with an infall velocity of about $c_{\rm cl}/3$ corresponds well to the results of the simulation of the formation of a primordial star by \citet{abel02}.  \citet{tm04} adopted this solution, noting that it has an accretion rate that is $2.6$ times greater than that for a static initial condition \citep{shu77} when expressed in dimensionless form.  Our problem is quite different from Hunter's, since an equilibrium density gradient $\krho = 1$ corresponds to $\gamma = 0$ \citep{mt03} rather than Hunter's $\gamma = 1$.  Nonetheless, we assume that the accretion rate for our problem is also 2.6 times greater than that for a static initial condition and find
\begin{equation}
\dot{M}_{\rm acc,TM04} =  10.9\,G^2 \Sigma_{\rm res}^3 t^3.
\label{mdotacctm04}
\end{equation}
Clearly, this result is uncertain, and magnetic fields introduce further uncertainty.  Fortunately, we find that varying the numerical coefficient in equation~\ref{mdotacctm04} does not affect the qualitative nature of the results discussed below.

To model the effect of a finite gas supply, once the total mass of gas that has fallen onto the cloud exceeds the total mass of the reservoir, $M_{\rm res}$, we set $\dot{M}_{\rm acc} = 0$.  When comparing with galactic populations of GMCs, we set $M_{\rm res} = 6 \times 10^6\unit{M}_\odot$,  the observed upper mass cutoff for GMCs \citep{wm97,fk10}.  This may underestimate the true upper mass since fragmentation may lead to a range of reservoir masses.  Since $6 \times 10^6\unit{M}_\odot$ GMCs are relatively rare, we set $M_{\rm res} = 2\times10^6 \unit{M}_\odot$ for the runs presented in Figure~\ref{cloudpropsfeedback} and Table~\ref{runoutcomes} as $\sim 10^6 \unit{M}_\odot$ is a more typical GMC mass.

If we also assume that the atomic reservoir is virialized such that the virial parameter of the reservoir, $\alpha_{\rm res}$, is constant with radius, we find
\begin{equation}
\begin{split}
\sigma_{\rm res} &= 0.4 \alpha_{\rm res}^{1/2} G \Sigma_{\rm res} t
\label{sigmares}
\end{split}
\end{equation}
Fiducially, we take $\alpha_{\rm res} = 2.0$, corresponding to a marginally gravitationally bound reservoir.  This parameterization assumes that the reservoir satisfies an internal linewidth-size relation of the form $\sigma_{\rm res}(r) \propto r^{1/2}$.  The increase of the reservoir velocity dispersion with time reflects that material originates at increasingly larger radii.

The precise normalization of the mass accretion law is a major source of uncertainty in our modeling.  Since the length scales over which material is swept up into the cloud through the reservoir approach galactic dynamical scales \citep{dob11,tas11}, a more complete treatment would require tracking the gas dynamics from the scale of an entire galaxy down to the the scale of the reservoir. This would also allow us to self-consistently model the end of accretion onto the cloud instead of assuming that mass accretion cuts off abruptly.  In a forthcoming paper, we plan to add our molecular cloud models to a simulation of gas dynamics in a galactic disk to model the gas reservoir for a population of GMCs.

Although the normalization of the accretion law is somewhat uncertain, we can make use of observations of the gas content of nearby galaxies to estimate $\Sigma_{\rm res}$, the surface density of gas in the reservoir feeding the cloud.  The average {\hi} surface density in the inner disk of the Milky Way is observed to be approximately constant, $\sim 8 \unit{M}_\odot \unit{pc}^{-2}$ \citep{kal08}.  Beyond a galactocentric radius of $\sim 10$ -- $15 \unit{kpc}$, the {\hi} surface density exponentially decreases with radius, however, few GMCs are observed in the outer Milky Way \citep{hey01} or beyond the optical radius of nearby galaxies \citep{eng03,big08}.  Similar saturated mean {\hi} surface densities are observed in nearby galaxies, except in the central regions of some galaxies where the ISM becomes fully molecular and the {\hi} surface density goes to zero \citep{ler08}.  For this reason, we adopt $\Sigma_{\rm res} = 8 \unit{M}_\odot \unit{pc}^{-2}$ as a fiducial atomic reservoir surface density typical of the bulk ISM of local star forming galaxies.

Although the {\it mean} atomic surface density in nearby galaxies is as low as $8 \unit{M}_\odot \unit{pc}^{-2}$, the atomic ISM is observed to be clumpy, with overdense regions reaching significantly higher surface densities.  These regions may be associated with spiral arms, as in M 33 \citep{thil02}, or driven by gravitational instability, as in the LMC \citep{yang07}.  For this reason, we also explore the behavior of molecular clouds accreting from higher surface density gas, $\Sigma_{\rm res} = 16 \unit{M}_\odot \unit{pc}^{-2}$.  Since massive molecular clouds are universally observed to be associated with high surface density gas \citep{won09,imara11}, we expect there to be marked differences between molecular clouds that accrete from high surface density gas and clouds that accrete from low surface density gas.  Although the gas will be primarily atomic at a gas surface density of $16 \unit{M}_\odot \unit{pc}^{-2}$, we expect that there should be some diffuse ``dark" molecular gas \citep{kmt08,wol10}.  Thus the reservoir is not necessarily completely atomic, but instead primarily composed of atomic gas.

\subsection{Numerical Scheme}
\label{numericalscheme}
Equations~(\ref{mdot_def}),~(\ref{reqn}), and~(\ref{sigmaeqn}) constitute a system of coupled, stochastic, nonlinear ordinary differential equations in $M$, $\sigma$, and $R$.  We solve these equations by using a straightforward Euler integration with an adaptive step size.  The precise order in which we update cloud properties is as follows.  After calculating the instantaneous star formation rate, we calculate $M^\prime$ by summing the components due to ionized winds and mass accretion.  Next, we calculate the rate of turbulent dissipation using equation~(\ref{turbdecay}).  We then calculate $\zeta$, the ratio of the mass doubling time to the free-fall time, using equation~(\ref{vineqn}).  We use $\zeta$ to calculate $a^\prime$, $f$, and $\xi$ by interpolating on precomputed tables.  Since $\sigma^\prime$ depends on $R^{\prime\prime}$, we first evaluate $R^{\prime\prime}$ using equation~(\ref{reqn}) and then compute $\sigma^\prime$ using equation~(\ref{sigmaeqn}).  Next, we check if $R$, $M$, or $\sigma$ will change by more that $0.1 \,\%$ using the current value of the time step.  If we detect a change larger than this, the time step is iteratively recalculated using a new time step half the size of the original until the fractional changes in $R$, $M$, and $\sigma$ are smaller than $0.1 \,\%$.   Next, we calculate $R$, $M$, and $\sigma$ at the new time step, update the state of any H~\textsc{ii} regions created in previous time steps, and then create new H~\textsc{ii} regions using the procedure described in \S\,\ref{starformation}.  If the time step did not need to be reduced, we increase the size of the time step by $10\%$.

Cloud evolution can be terminated if one of three conditions is satisfied:  
\begin{itemize}
\item The time step is less than $10^{-8}$ of the current evolution time (i.e.\ $\Delta\tau/\tau < 10^{-8}$).
\item The mean visual extinction falls below $A_{V\rm, min} = 1.4$, corresponding to the CO dissociation threshold found by \citet{van88}.
\item An H~\textsc{ii} region envelops and unbinds the cloud, i.e. if $r_{\rm sh} > R_{\rm cl}$ and $\dot{r}_{\rm sh} > v_{\rm esc}$.  
\end{itemize}
We use the phrases collapse, dissociation, and disruption, respectively to describe these scenarios.  The dissociation threshold depends on the ambient radiation field.  However, \citet{wol10} found that the CO dissociation threshold varies by only a factor of two when the intensity of the ambient radiation field varies by an order of magnitude, so we neglect variations in the radiation field and for consistency with Paper I, adopt $A_{V,\rm min} = 1.4$.  The surface density corresponding to the dissociation threshold depends on the assumed dust to gas ratio and thus on the metallicity.  For solar metallicity, a surface density of $1 \unit{g} \unit{cm}^{-2}$ corresponds to $A_V = 214.3$.  Since we use a dissociation threshold based on CO rather than H$_2$ to define the end of the cloud's life, we may miss the further evolution of a diffuse molecular cloud where most of the carbon is neutral or singly ionized but the hydrogen is still molecular.

These halting conditions probably oversimplify the true end of a cloud's evolution due to our assumption of spherical symmetry and homology.  For the case of collapse, it is more likely that the cloud would undergo runaway fragmentation rather than monolithic collapse.  In the case of dissociation, even if the mean surface density drops below the point where CO can no longer remain molecular, that does not preclude the possibility that overdense clumps might retain significant amounts of CO.  Finally, for the case of disruption, even if an H~\textsc{ii} region delivered a large enough impulse to unbind the cloud, it may simply be displaced as a whole, or be disrupted into multiple pieces which would then evolve independently.  It is also likely that if the cloud is disrupted while still actively accreting, the cloud would inevitably recollapse since it is unlikely that an H~\textsc{ii} region would have enough kinetic energy to unbind the reservoir.

Since we must necessarily use simple criteria to halt the cloud evolution, our estimates of cloud lifetimes presented below are strictly lower limits to the true lifetime of a cloud.  Both the disruption and dissociation criteria do not preclude the presence of overdense clumps that may survive the destruction events.  However, our estimates of cloud lifetimes are appropriate for the lifetime of a single monolithic cloud.  Any overdense clumps that do survive would represent entirely different clouds that would evolve independently of each other.

\subsection{Input Parameters}

\begin{deluxetable}{cccc}
\tablecaption{Fiducial Parameters}
\tabletypesize{\scriptsize}
\tablewidth{0pt}
\tablehead{
\colhead{Parameter} & \colhead{Value} & \colhead{Reference}
}
\startdata
$\alpha_{\rm vir,0}$ & 2.0 &  \citet{bli07}\\
$\Sigma_{\rm cl,0}$ & $60 \unit{M}_\odot \unit{pc}^{-2}$ &  \citet{hey09}\\
$c_s$\tablenotemark{a} & $0.19 \unit{km}\unit{s}^{-1}$ & ---\\
$c_\textsc{ii}$\tablenotemark{b} & $9.74 \unit{km} \unit{s}^{-1}$ &  \citet{mw97}\\
$\eta_B$ & 0.5 &  \citet{km05}\\
$\eta_E$ & 1.0 &  Paper I\\
$\eta_v$ & 1.2 &  Paper I\\
$\phi_{\rm in}$ & 1.0 &  \citet{bru09}\\
$\varphi$ & 0.75 &  This work\\
$A_{\rm v,min}$& $1.4 \unit{M}_\odot \unit{pc}^{-2}$ & \citet{van88}\\
$\alpha_{\rm res}$ & 2.0 & ---\\
$P_{\rm amb}/k_{\rm B}$ & $3\times 10^{4} \unit{K}\unit{cm}^{-3}$ &  \citet{mck99} \\
$M_{\rm cl,0}$ & $5 \times 10^4 \unit{M}_\odot$ & ---
\enddata
\tablenotetext{a}{Assumes $T = 10 \unit{K}$ and $\mu = 2.3$}
\tablenotetext{b}{Assumes $T = 7000 \unit{K}$ and $\mu = 0.61$}
\label{cloudparamtable}
\end{deluxetable}

To complete our model, we must choose a set of parameters and initial conditions to fully determine our cloud evolution equations.  In Table~\ref{cloudparamtable} we have listed the various fiducial parameters we have chosen for our model as well as the references from which we derive our choices.  Some of these parameters are motivated by observations, others by the results of simulations, and one parameter ($\varphi$, see \S\,\ref{varphivary}) is left free.

The cloud initial conditions can be computed given an initial cloud mass along with our assumed value for $\alpha_{\rm vir,0}$ and $\Sigma_{\rm cl,0}$.  \citet{mat00} and \citet{fal10} found that protostellar outflows are energetically important when $M_{\rm cl} \lesssim 10^{4.5} \unit{M}_\odot$.  Above this mass, they contribute negligibly.  Thus, if we choose a low initial mass, we may be underestimating the amount of turbulent energy injection by star formation feedback at early times since we do not account for protostellar outflows.  For this reason, we choose a relatively large initial mass, $M_{\rm cl,0} = 5\times10^4 \unit{M}_\odot$.  For reference, given our choice of initial virial parameter and surface density, this corresponds to, $R_{\rm cl,0} = 11.5 \unit{pc}$, $\sigma_{\rm cl,0} = 2.7 \unit{km} \unit{s}^{-1}$, and $t_{\rm cr,0} = 4.1 \unit{Myr}$.  While this is larger than some local molecular clouds like Taurus or Perseus, it is still much smaller than the mass of the molecular clouds where most of the star-formation in local galaxies occurs.  This choice also ensures that we are accounting for the bulk of the energy available from star formation feedback due to photoionization, ionized gas pressure, and radiation pressure.

\section{Models with Accretion Only}
\label{varphivary}

Before beginning full simulations using our method, we must first test the behavior of our model as we vary the free parameter $\varphi$ introduced in the derivation of the energy evolution equation. For this purpose, we have run our model with star formation feedback disabled.  The only physical mechanisms modeled in these tests are accretion and the decay of turbulence. Since there is no random drawing from the stellar or cluster IMF in these runs, the results are fully deterministic. These simplified models allow us to understand how the results depend on our choice for the tunable parameter $\varphi$ and provide physical insight that will be useful in interpreting the results of the more complex and stochastic runs that include feedback.

The energetics and virial balance of our cloud models depend critically on the parameter $\varphi$.  Broadly speaking, $\varphi$ controls the amount of turbulent kinetic energy injected by the accretion flow.  For the case $\varphi = 0$ the accretion flow contributes the minimum possible amount of turbulent kinetic energy. This means that accreted material cannot contribute significantly to turbulent pressure support, since the accreted material is maximally subvirial.  With this choice, as the cloud accretes mass, its energy budget must become increasingly dominated by self-gravity.  Once this happens, internal pressure support is negligible and the cloud must inevitably undergo gravitational collapse.

Alternatively, we could set $\varphi > 0$.  If $\varphi$ is small, the turbulent kinetic energy of a newly accreted parcel of gas would still be small compared to the gravitational potential energy of the gas parcel.  Thus, once the cloud is primarily composed accreted gas, the cloud will undergo gravitational collapse, although on a slightly longer timescale than in the $\varphi=0$ case.  At some larger value of $\varphi$, accretion contributes a net positive amount of energy to the cloud, balancing out the negative gravitational potential energy of the newly accreted material.  The could will still collapse with this choice since turbulent motions quickly decay away.  As $\varphi$ is increased further, we should eventually find that at some critical value, $\varphi = \varphi_{\rm crit}$, accretion drives turbulent motions with sufficient vigor to avoid the gravitational collapse of the cloud entirely.

\begin{figure}
\resizebox{\columnwidth}{!}{\includegraphics{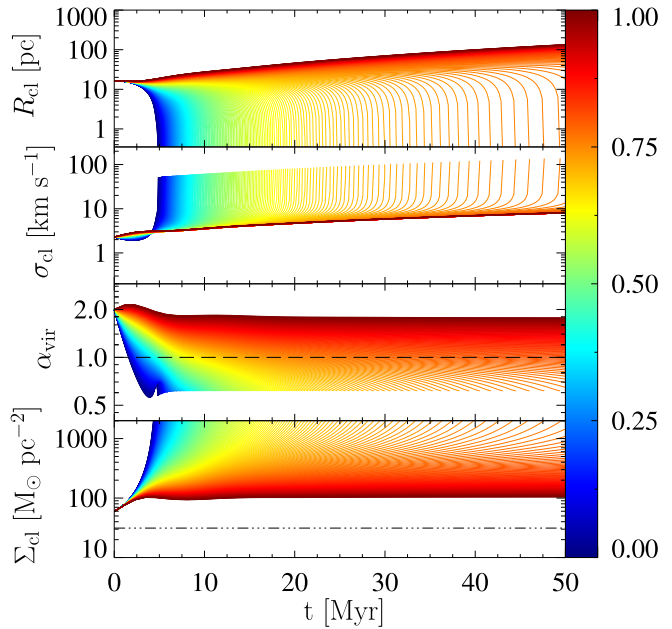}}
\caption{Cloud surface densities (\emph{bottom row}), virial parameters (\emph{second row}), velocity dispersions (\emph{third row}), and radii (\emph{top row}) for 400 different runs, each with a different choices for $\varphi$, as indicated in the color bar.  Star formation was turned off for all runs.}
\label{cloudpropsphicorr} 
\end{figure}

The results of test runs with different choices of $\varphi$ are presented in Figure~\ref{cloudpropsphicorr}.  The time evolution of the cloud surface density, virial parameter, velocity dispersion, and cloud radius is plotted for a selection of clouds evolved with different choices for $\varphi$.  Each line depicts the time evolution of a cloud property and is color coded by the value of $\varphi$ chosen.  It is obvious that the value of $\varphi$ can strongly influence the resulting evolution.  

If $\varphi = 0$, the cloud experiences global collapse in a free-fall time.  Initially, the cloud velocity dispersion decreases, but inevitably the gravitational term in the velocity dispersion evolution equation, proportional to $-R^\prime / R^2$, becomes dominant, and the velocity dispersion begins to diverge.  The fact that $\sigma_{\rm cl}$ diverges as $R_{\rm cl}$ goes to zero is an artifact.  In reality, the highest density regions would independently fragment and collapse and the cloud would never undergo a monolithic collapse.   

As we increase $\varphi$, the cloud is able to support itself against collapse for longer periods.  Near $\varphi = 0.5$, accretion brings in net positive energy but turbulent dissipation wins out, and the cloud still eventually collapses.  At a critical value, $\varphi_{\rm crit} \simeq 0.8$, accretion driven turbulence alone is sufficient to hold up the cloud against collapse for as long as the reservoir continues to supply mass to the cloud.  The mass, radius, and velocity dispersion of the cloud increase in such a way as to maintain a constant virial parameter and surface density.

Since we expect that gas motions driven by accreting dense clumps should be at least somewhat correlated with the motions of the infalling clumps, we do not expect a physically realistic choice of $\varphi$ to be very close to zero.  On the other hand, a model in which a cloud is entirely supported by accretion driven turbulence seems to preclude the possibility that a significant fraction of the kinetic energy of infalling gas is radiated away in an accretion shock.  For this reason, we rule out as unphysical runs with $\varphi \approx 0$ and $\varphi \ge \varphi_{\rm crit}$.  The precise value of $\varphi$ we will use in our models that include star formation below depends on uncertain details of the accretion and mixing of infalling gas.  In practice, we find that even with the energy provided by star formation feedback, clouds generally undergo free-fall collapse or reach unreasonably high mean surface densities once they are primarily composed of accreted material if we choose $\varphi \lesssim 0.7$.  Since clouds are generally not observed to be in global free-fall collapse, we instead pick a value somewhat higher that this, $\varphi = 0.75$ for our fiducial models.  This splits the difference between accretion contributing a negligible amount of energy to the cloud when $\varphi = 0.5$ and accretion contributing the maximum possible amount of energy when $\varphi = 1$.  We will see below that our fiducial choice broadly reproduces the observed properties of molecular clouds in the Milky Way and nearby galaxies.  

\section{Models with Accretion and Star Formation}
\label{fullsims}

Feedback by the action of ionizing radiation emitted by newborn stellar associations alters the evolution of a GMC after the birth of the first massive star cluster.  The source of energy provided by massive star formation can be a significant component of the energy budget of the entire cloud.  For the remainder of this paper, we consider models with the star formation prescription described in \S\,\ref{starformation} turned on.

\subsection{Overview of Results}

\begin{figure*}
\resizebox{\textwidth}{!}{\includegraphics{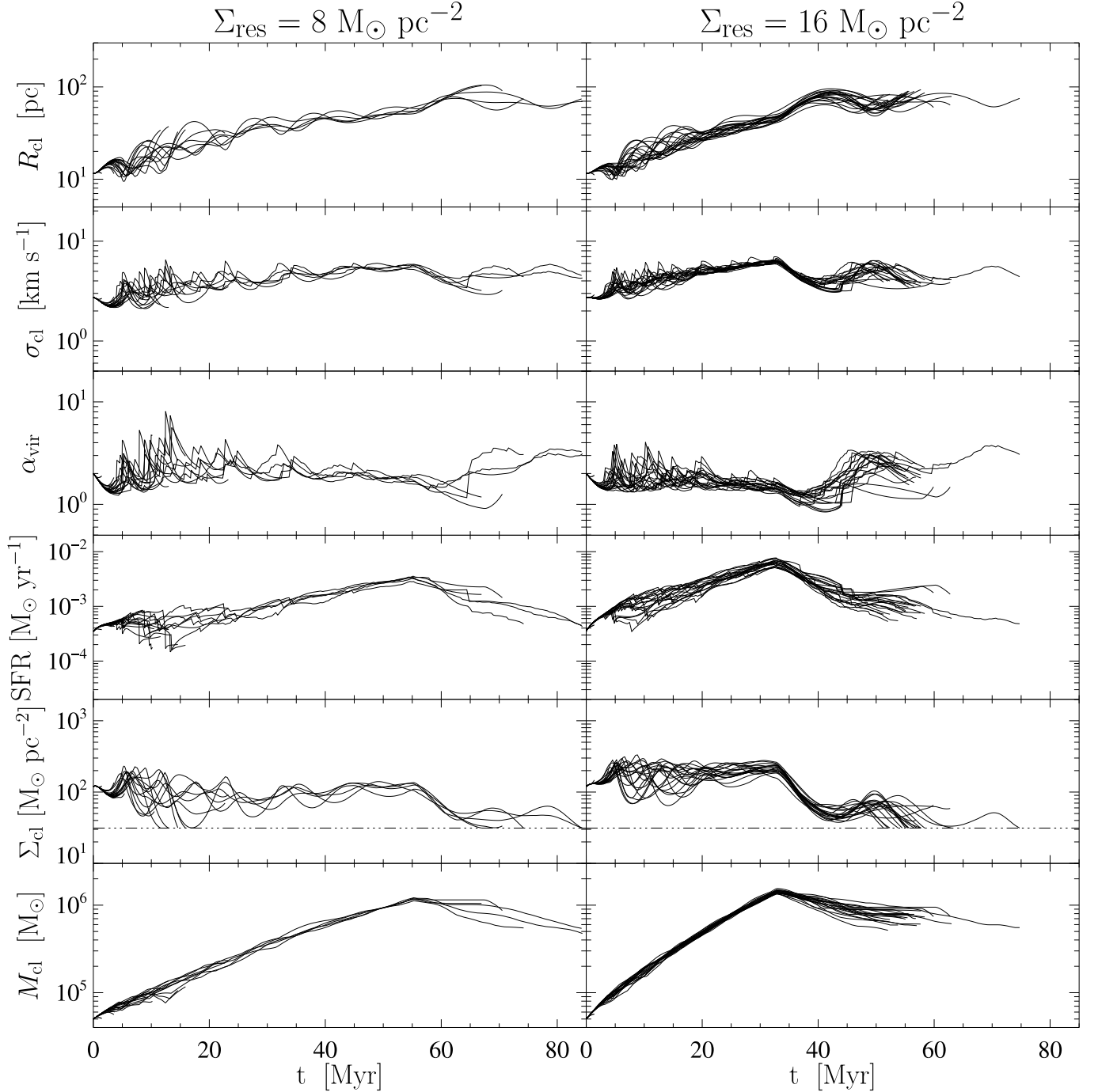}}
\caption{Cloud surface densities (\emph{bottom row}), star formation rates (\emph{second row}), virial parameters (\emph{third row}), velocity dispersions (\emph{fourth row}), and radii (\emph{top row}) for a set of 40 clouds.  Half of the clouds were evolved with a low surface density characteristic of the bulk of the atomic ISM and the other half were evolved with a high surface density, characteristic of overdense regions in the ISM.  The two different choices of $\Sigma_{\rm res}$ are marked at top.  Accretion was shut off once $10^6 \unit{M}_\odot$ of material had been processed through the accretion flow.  The ambient surface density, and thus the accretion history, strongly affects the resulting cloud evolution.}
\label{cloudpropsfeedback}
\end{figure*}

We have run two sets simulations with parameters chosen to model conditions in interarm ($\Sigma_{\rm res} = 8 \unit{M}_\odot \unit{pc}^{-2}$) and spiral arm ($\Sigma_{\rm res} = 16 \unit{M}_\odot \unit{pc}^{-2}$) regions.  Besides the two different choices for the ambient surface density, all other parameters and initial conditions are identical.  The time evolution of a subsample of runs are plotted in Figure~\ref{cloudpropsfeedback} and average properties of the full sample are presented in Table~\ref{runoutcomes}.

\begin{deluxetable*}{cccccccc}
\tablecaption{Average Properties of Model Clouds}
\tabletypesize{\scriptsize}
\tablewidth{0pt}
\tablehead{
\colhead{$\Sigma_{\rm res}$} & \colhead{$\avg{t_{\rm life}}$} & \colhead{$\avg{M_{\rm max}}$} &\colhead{$\avg{M_{\rm ejected}}$} & \colhead{$\avg{\epsilon}$} & \colhead{$\avg{\epsilon_{\rm ff}}$} & \colhead{$N_{\rm dissoc}$} & \colhead{$N_{\rm disrupt}$} \\[3pt]
\colhead{[${\rm M}_\odot \unit{pc}^{-2}]$} & \colhead{[Myr]} & \colhead{[${\rm M}_\odot$]} & \colhead{[${\rm M}_\odot$]} & \colhead{[$\%$]} & \colhead{[$\%$]} & &
}
\startdata
8 & $26.2 \pm 29.8$ & $(3.7 \pm 4.9) \times 10^5$ & $(3.4 \pm 5.3) \times 10^5$ & $5.0 \pm 2.3$ & $1.9 \pm 0.5$ & 308 & 692\\
16 & $52.6 \pm 16.8$ & $(1.3 \pm 0.4) \times 10^6$ & $(1.2 \pm 0.4) \times 10^6$ & $8.3 \pm 2.0$ & $1.9 \pm 0.4$ & 687 & 313
\enddata
\label{runoutcomes}
\end{deluxetable*}

The most striking result of our comparison is that the final mass of our model molecular clouds depends on the assumed mass accretion history.  Clouds evolved with a low accretion rate, corresponding to conditions in interarm regions, grow larger than $10^{5} \unit{M}_\odot$ less than $30 \%$ of the time and very rarely reach masses comparable to the most massive GMCs in the local group.  The vast majority of clouds are instead disrupted by an energetic {\hii} region within a few crossing times.  The clouds attain a quasi-equilibrium configuration in which mass accretion is roughly balanced by mass ejection.  Clouds avoid global collapse by extracting energy from the expansion of {\hii} regions.  

The evolution of the clouds is characterized by discrete energy injection events due to the formation of a single massive star cluster.  Once a cluster forms, it ejects a wind and launches an {\hii} region.  The recoil force of launching the wind leads to an overall confining ram pressure, causing the radius to decrease and the surface density to increase.  Once the star cluster burns out, the {\hii} region expansion decelerates and then stalls.  When the expansion velocity of the {\hii} region is comparable to the cloud velocity dispersion, the kinetic energy of the expanding {\hii} region is converted into turbulent kinetic energy, causing a spike in the turbulent velocity dispersion.  The turbulent kinetic energy exponentially decays away over a crossing time, but the temporarily elevated velocity dispersion increases the turbulent kinetic pressure, causing the cloud to expand.  This leads to oscillations in the cloud radius and mean surface density.  On the whole, clouds that are not quickly disrupted by {\hii} regions, are able to survive as quasi-virialized objects for several crossing times before they are either disrupted or dissociated.

Clouds evolved with a higher ambient surface density, typical of spiral arm regions in the Milky Way, exhibit significantly different behavior.  Since these clouds accrete mass much faster than in the low surface density runs, they are not able to attain steady state between accretion and ejection of mass.  While some clouds are still destroyed by energetic {\hii} regions early in their evolution, over $90 \%$ of these clouds were able to accrete their entire reservoir after $25 \unit{Myr}$.  At this point, the clouds are generally quite massive, $\sim 1.5\times10^6 \unit{M}_\odot$.   Once accretion is shut off, the clouds are no longer confined by accretion ram pressure and lose a portion of the power that had been driving turbulence.  For this reason, the velocity dispersion decreases in response to the loss of accretion driven turbulence, and the cloud radius expands in response to the loss of the confining pressure provided by accretion.  Before the cloud can dissociate, it attains pressure balance with the ambient ISM at a lower velocity dispersion and larger radius.  For the next 20 to 30 Myr,  the clouds evolve in much the same way as the massive cloud models considered in Paper I.  The clouds can be supported against self-gravity for many dynamical times by forming stars and launching {\hii} regions.  Particularly energetic {\hii} regions can disrupt the clouds and excursions to low surface density can dissociate the clouds.  The lifetime of these clouds is thus set by the amount of time they can accrete.  This may imply that spiral arm passage times set GMC lifetimes, although further work is needed to clarify this tentative conclusion.

The star forming properties of the two sets of models are also somewhat different.  In the interarm case, the star formation efficiency,
\begin{equation}
\epsilon = \frac{M_{*,{\rm tot}}}{\int_0^{t_{\rm life}}\dot{M}_{\rm acc}dt},
\label{sfe}
\end{equation}

\noindent
is only $5\%$ while in the high surface density case, $\epsilon$ is somewhat larger, approximately $8 \%$.  This can be entirely attributed to the difference in lifetimes between the two sets of models.  In the low surface density case, most clouds are only able to survive one or two crossing times and thus can only convert a small fraction of their mass into stars before they are dissociated or disrupted.  The clouds evolved with a high ambient surface density are able to survive for many crossing times and convert a larger fraction of their gas into stars.  An even larger fraction is ejected via photoionization.  However, for both models, the star formation efficiency per free-fall time,  
\begin{equation}
\epsilon_{\rm ff} = \frac{\dot{M}_{*}}{M_{\rm cl}t_{\rm ff}}
\label{sfeff}
\end{equation}  

\noindent
is low, around $2\%$.  This is not surprising, as a low star formation efficiency per free-fall time is one of the basic assumptions of our model.

\subsection{Energetics of Star Formation Feedback Versus Mass Accretion}
\label{energetics}

GMCs exhibit highly supersonic turbulence.  There is no agreement in the literature about what drives these motions, which numerical models of compressible MHD turbulence indicate should decay if left undriven.  Some authors suggest that the primary energy injection mechanism is some sort of internal star formation feedback process, such as protostellar outflows \citep{lin06,wan10}, expanding H~\textsc{ii} regions \citep{mat02}, or supernovae \citep{mlk04}.  Others suggest that turbulence is driven externally via mass inflows \citep{kh10}.  Comparing the amount of energy injected by different forms of star formation feedback, \citet{fal10} found that at typical GMC column densities, the dominant stellar feedback mechanism is H~\textsc{ii} regions driven by the intense radiation fields emitted by massive star clusters.  Using our models, we can compare the importance of accretion relative to H~\textsc{ii} regions in the energy budget of GMCs.

To find the total energy injected by accretion, we make use of our knowledge of the total energy of the cloud as a function of time. At the end of time step $j$ we use  equation~(\ref{cloudenergy}) to calculate both the total cloud energy, $\mathcal{E}_{{\rm cl},j}$, as well as what the cloud energy would have been if we had set $\dot{M}_{\rm acc} = 0$ for that time step, $\mathcal{E}_{\rm cl}|_{\dot{M}_{\rm acc} = 0}$. The difference, 
\begin{equation}
\mathcal{E}_{\rm acc,j} = \mathcal{E}_{{\rm cl},j} - \mathcal{E}_{\rm cl}|_{\dot{M}_{\rm acc} = 0}
\label{eaccjdef}
\end{equation}
is the total energy added by accretion during that time step.  The total energy injected by accretion over the cloud's lifetime is just the sum of the contributions of each time step,
\begin{equation}
\mathcal{E}_{\rm acc} = \sum_j \mathcal{E}_{{\rm acc,}j}.
\end{equation}

The energy injected by H~\textsc{ii} region $i$, $\mathcal{E}_{{\rm H\,\textsc{ii},}i}$, can be found by integrating the rate of energy injection by a single H~\textsc{ii} region with respect to time.  This is,
\begin{equation}
\mathcal{E}_{{\rm H\, \textsc{ii}},i} = 1.6\eta_E\mathcal{T}_{1,i}\left(\frac{r_{{\rm m,}i}}{R_{{\rm cl},i}}\right)^{1/2}
\end{equation}
where $r_{{\rm m},i}$ is the radius of H~\textsc{ii} region $i$ when it merges with the parent cloud and $R_{{\rm cl},i}$ is the radius of the cloud as a whole when {\hii} region $i$ merged with the cloud.  To find the total energy injected by H~\textsc{ii} regions over the cloud's lifetime, we simply sum up the contributions due to individual H~\textsc{ii} regions,
\begin{equation}
\mathcal{E}_{\rm H\, \textsc{ii}} = \sum_i \mathcal{E}_{{\rm H\,\textsc{ii}, }i}.
\end{equation}

The ratio $\abs{\mathcal{E}_{\rm H~\textsc{ii}} / \mathcal{E}_{\rm acc}}$ indicates the relative importance of star formation feedback to accretion driven turbulence to the global energy budget of the cloud.  If $\abs{\mathcal{E}_{\rm H~\textsc{ii}} / \mathcal{E}_{\rm acc}} < 1$, accretion dominates the energy injection; similarly if $\abs{\mathcal{E}_{\rm H~\textsc{ii}} / \mathcal{E}_{\rm acc}} > 1$, star formation feedback is the primary driver of turbulence.

\begin{figure*}
\resizebox{\textwidth}{!}{\includegraphics{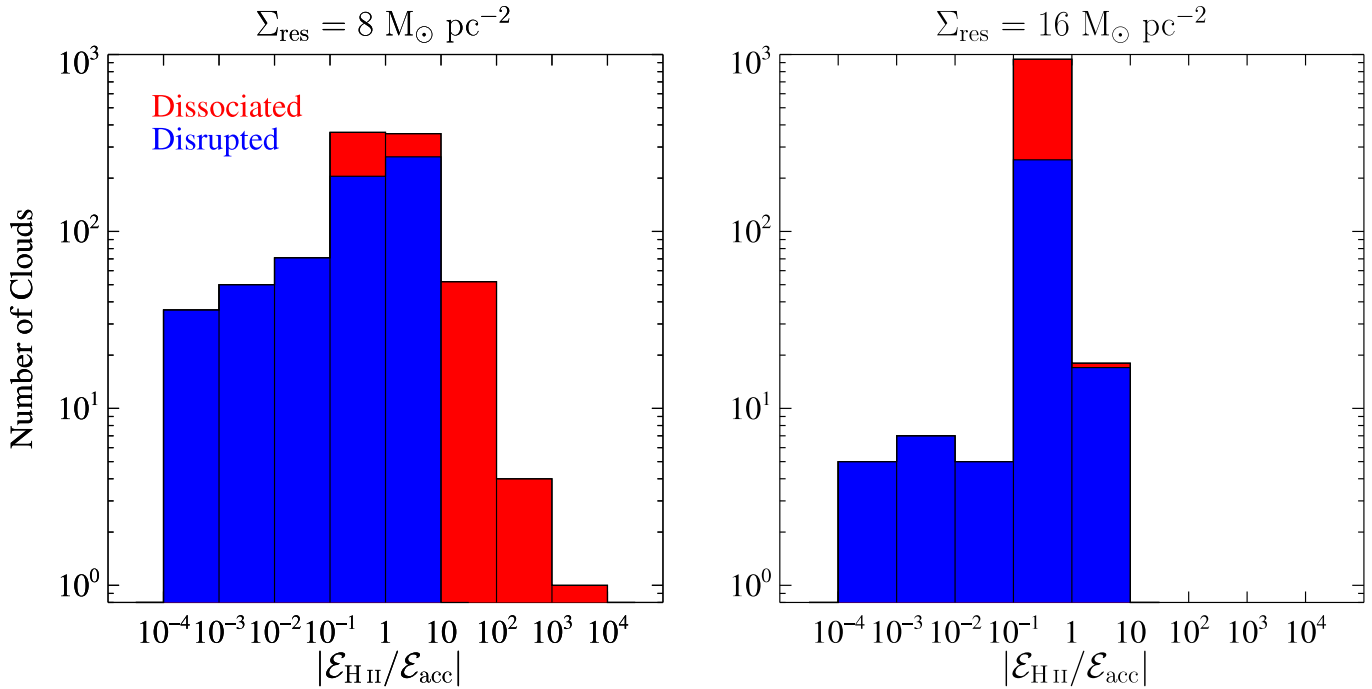}}
\caption{The number of clouds plotted as a function of  $\abs{\mathcal{E}_{\rm H~\textsc{ii}} / \mathcal{E}_{\rm acc}}$.  In regions of low ambient surface density, accretion and star formation are in equipartition, while in regions of high ambient surface density, accretion dominates the energy budget.}
\label{faccMdotPlot}
\end{figure*}

The results of this comparison are plotted for both choices of the ambient surface density in Figure~\ref{faccMdotPlot}.  We find that {\hii} regions and accretion contribute approximately equal amounts of energy in the low surface density runs, while accretion dominates in the high surface density runs.  In the low surface density runs, stochastic effects can be important, particularly for clouds that do not last much longer than a crossing time.  Thus, in some runs, star formation feedback can contribute significantly more energy than accretion, while in others star formation feedback is negligible.  In the runs evolved with a high ambient surface density, star formation feedback is subdominant, although not completely negligible, in the vast majority of runs.

It is worth pointing out that this result depends on the precise value of $\varphi$ we choose to evolve the clouds with.  If $\varphi$ is lower, accretion contributes less energy, and star formation can dominate the energy budget.  If $\varphi$ is higher, star formation becomes completely negligible, and the amount of kinetic energy injection is controlled by the mass accretion rate.  Since clouds collapse when we choose $\varphi$ much lower than our fiducial value, and shocks in molecular gas tend to be strongly dissipative, we do not expect the `true' value of $\varphi$ to be much different than our fiducial value.  We thus conclude that one of three cases must hold.  Star formation may be dominant, but only marginally so.  Accretion may also be dominant, but again, only marginally.  It is also possible that star formation and accretion contribute roughly equal amounts of energy.  In all three cases, neither star formation or accretion is truly negligible.

\section{Observational Comparisons}
\label{obscompare}

\subsection{Larson's Laws}
Giant molecular clouds are observed to obey three scaling relations, known as Larson's laws \citep{lar81,sol87,bol08}.  In their simplest form, Larson's laws state:

\begin{itemize}
\item The velocity dispersion scales with a power of the size of the cloud.  Subsequent observations have shown that this power is about $0.5$ ($\sigma_{\rm cl} \propto R_{\rm cl}^{0.5}$).
\item The mass of the cloud scales with the square of the radius (constant $\Sigma_{\rm cl}$).
\item Clouds are in approximate virial equilibrium ($\alpha_{\rm vir}$ of order unity).
\end{itemize}

These laws are not independent; any two imply the other.  At a minimum, an acceptable theoretical model for GMCs should agree with both the scaling and the normalization of the Larson scaling relations observed in real clouds.  We have already seen that clouds maintain approximate virial equilibrium as well as roughly constant surface densities, but we have yet to see whether the normalization of the Larson scaling relations for our models agrees with the observed Larson scaling relations.

\subsubsection{Equilibrium Surface Densities}

GMCs, both in the Milky Way \citep{lar81,sol87}, and in nearby external galaxies \citep{bli07,bol08} exhibit surprisingly little variation in surface density.  For the \citet{sol87} sample of Milky Way clouds, this was found to be $\left\langle \Sigma_{\rm cl} \right\rangle = 170 \unit{M}_\odot \unit{pc}^{-2}$.  More recent and sensitive observations find lower values, closer to $\left\langle \Sigma_{\rm cl} \right\rangle = 50 \unit{M}_\odot \unit{pc}^{-2}$ in the Milky Way \citep{hey09} and in the Large Magellanic Cloud \citep{hug10}, although these latter estimates depend on a highly uncertain correction for non-LTE line excitation and the CO to $\textrm{H}_2$ conversion factor, respectively.  Using heterogenous data from several nearby galaxies, \citet{bol08} attempted to extract cloud properties in a uniform manner and found a typical surface density of $85 \unit{M}_\odot\unit{pc}^{-2}$ but with significant variation from galaxy to galaxy.  

Variations  in the mean GMC surface density are seen when comparing samples from different galaxies.  However, within a single galaxy there is little variation \citep{bli07}.  These variations are usually attributed to differences in the CO-to-$\rm{H}_2$ conversion factor from galaxy to galaxy \citep{bol08}, a quantity which may depend on metallicity and the interstellar radiation field \citep{gm10} as well as variations in turbulent pressure and radiation field in the ambient interstellar medium.  In our runs, we also recover roughly constant surface densities (see the second row from the bottom of Figure~\ref{cloudpropsfeedback}).

\begin{figure}
\resizebox{\columnwidth}{!}{\includegraphics{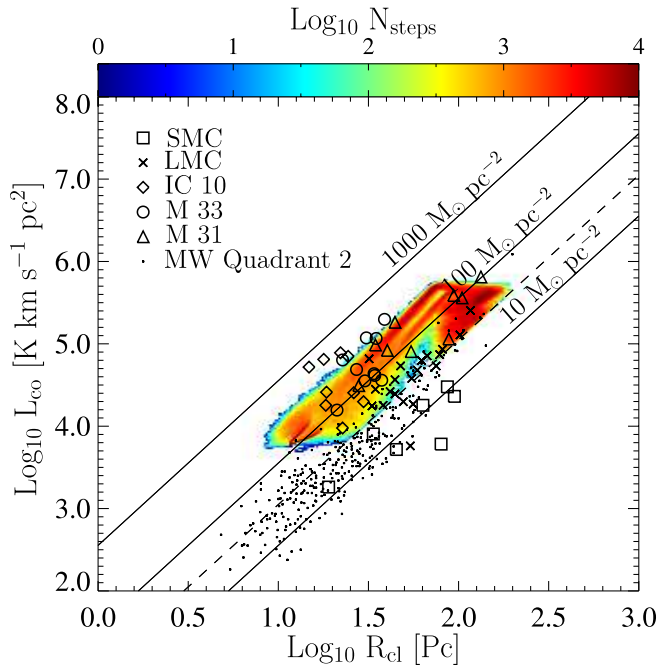}}
\caption{Cloud CO luminosity plotted as a function of cloud radius.  CO luminosities are found by assuming $X_{\rm CO} = 4 \times 10^{20} \unit{cm}^{-2} (\unit{K} \unit{km} \unit{s}^{-1})^{-1}$.  Solid lines of constant surface density are plotted for $10$, $100$, and $1000 \unit{M}_\odot \unit{pc}^{-2}$ for reference.  The dashed line of constant surface density corresponds to our assumed dissociation threshold.  The outputs from a set of 2000 runs were used, with $\Sigma_{\rm res} = 8 \textrm{ and } 16 \unit{M}_\odot \unit{pc}^{-2}$ and $M_{\rm res} = 6\times10^6\unit{M}_\odot$.  Colors indicate the amount of time model clouds tend to occupy a position in $L_{\rm CO}$ -- $R_{\rm cl}$ parameter space.  Symbols denote observed CO luminosities and cloud radii for galactic (\emph{points}) and extragalactic (\emph{open shapes}) GMCs.  See \citet{bli07} and references therein for details of the observations.}
\label{blitzplot}
\end{figure}

In Figure~\ref{blitzplot}, we have reproduced a figure from \citet{bli07} that depicts observational results for CO luminosities and cloud radii for a sample of clouds in the outer Milky Way as well as from several samples of extragalactic GMCs.  To compare against this compendium of results, we calculate CO luminosities for our model clouds by assuming a constant CO-to-$\text{H}_2$ conversion factor,
\begin{equation}
L_{\rm CO} = \frac{M_{\rm cl}}{8.8 \unit{M}_\odot} \unit{K}\unit{km}\unit{s}^{-1}\unit{pc}^{2}
\end{equation}
as in \citet{rl06}.  This formula accounts for the presence of helium and assumes a constant $\rm{H}_2$ to CO conversion factor, $X_{\rm CO} = 4 \times 10^{20} \unit{cm}^{-2}\, (\unit{K}\unit{km}\unit{s}^{-1})^{-1}$, twice the value derived for molecular clouds within the Solar circle using observations of gamma-ray emission \citep{sm96,abd10}.   We choose this value to be consistent with \citet{bli07}, who find, using this value of $X_{\rm CO}$, that all of the GMCs in their sample have virial masses comparable to the masses implied by their CO luminosity to within a factor of two.
 
With our fiducial initial conditions, model clouds in our sample begin their lives in the bottom left hand corner of Figure~\ref{blitzplot}, at $R_{\rm cl} \approx 10 \unit{pc}$.  As they accrete and expand, clouds move towards the upper right hand corner.  Clouds end their evolution either through disruption by a single H~\textsc{ii} region or by passing below the molecular dissociation threshold, indicated by a dashed line in Figure~\ref{blitzplot}.  Offsets in the distribution of column densities from galaxy to galaxy and from the simulated clouds can be attributed to variations in $X_{\rm CO}$ and uncertainty in identifying a unique radius for observed clouds \citep{bli07} that have nonzero obliquity \citep{ber92}.  Accounting for variations in $X_{\rm CO}$, there is striking agreement between the observed distribution of molecular clouds and our sample of simulated clouds.

The models exhibit a kink in their evolution when the reservoir is exhausted and accretion is shut off.  For this reason, there are no clouds with $L_{\rm CO} > 10^{5.6} \unit{K}\unit{km}\unit{s}^{-1}\unit{pc}^2$.  Once accretion is shut off, the clouds decrease in mass for the remainder of their evolution.  This kink is somewhat artificial since we have assumed a fixed reservoir mass and a smooth accretion history.  A more sophisticated model for the reservoir including a range of reservoir masses would exhibit a continuous spectrum of kinks, broadening the region of parameter space explored by the models, particularly for $L_{\rm CO} < 10^{4.5} \unit{K}\unit{km}\unit{s}^{-1}\unit{pc}^2$.

The models also exhibit two distinct favored strips of parameter space along which they tend to evolve.  This corresponds to the two different equilibrium column densities picked out by the two different choices of $\Sigma_{\rm res}$.  This behavior is clearly seen in the second panel from the bottom of Figure~\ref{cloudpropsfeedback}.  The fact that $\Sigma_{\rm cl}$ is sensitive to $\Sigma_{\rm res}$ follows from dimensional analysis.

\subsubsection{Linewidth-Size Relation}
\begin{figure}
\resizebox{\columnwidth}{!}{\includegraphics{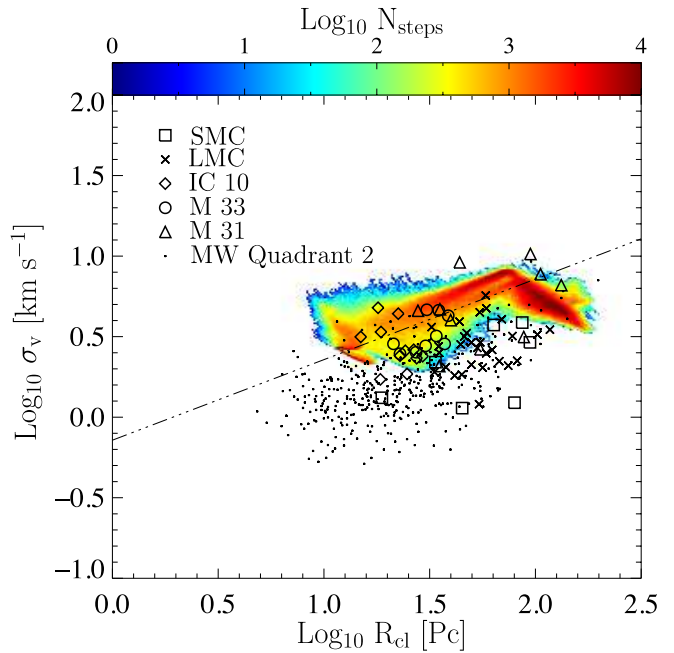}}
\caption{Cloud velocity dispersion plotted as a function of cloud radius.  The dash-dotted line is the galactic linewidth-size relation found by \citet{sol87}, $\sigma_{\rm v} = 0.72 R^{0.5}$.  Symbols and color coding are the same as in Figure~\ref{blitzplot}.}
\label{blitzplot_lw}
\end{figure}

We next compare our simulated clouds with the linewidth-size relation observed to hold among GMCs as a population \citep{bol08}.  We are able to reproduce the power law, scatter, and the rough normalization in the observed linewidth-size relation.  This conclusion is unsurprising, since we have already seen that our simulated clouds maintain roughly constant virial parameters and surface densities as they evolve.  It is worth noting that for our simplified model for the environment of a GMC, that the linewidth-size relation corresponds to an age sequence.  Clouds that live towards the left-hand side of the diagram are younger than clouds that live towards the right.  It is possible that this conclusion is an artifact of choosing a single reservoir mass.  Clouds accreting from a population of reservoirs with a continuous spectrum of masses may blur this effect somewhat.  We plan to revisit this in future work in which we will model the global ISM of a galaxy simultaneously with the evolution of a population of GMCs.

There is a small offset when comparing the locus of extragalactic and outer Milky Way clouds with our models, although there is good agreement between our models and the scaling found by \citet{sol87}.  For a subset of the observational sample, particularly the SMC clouds, it is possible that the metallicity of the gas in the clouds is so low that CO is no longer a good tracer of the bulk of the molecular gas \citep{ler07}.   Since our models assume perfect sphericity, and the observed radius of a prolate or oblate spheroid will always be smaller than the corresponding spherical radius \citep{ber92}, it is also possible that the radii predicted by our models overpredict the corresponding observed cloud radius by 0.1 or 0.2 dex.  Lastly, it could be the that we overpredict the various pressures due to photoionization and accretion by assuming spherical symmetry.  In reality, the wind and accretion ram pressure may not necessarily be perfectly spherically symmetric, leading to a reduction in the overall confining pressure and an increase in the radius.

\subsection{Evolutionary Classification}
\label{evocompare}

The Large Magellanic Cloud is home to one of the best-studied samples of GMCs in any galaxy.  The LMC's disk-like geometry and face-on orientation offers little ambiguity in distance measurements, with the most accurate measurements giving $d_{\rm LMC} = 50.1 \unit{kpc}$ \citep{alv04}.  A large quantity of high-quality multiwavelength data has been obtained for the entire disk of the galaxy.   In particular, the NANTEN $^{12}$CO $(J = 1 \rightarrow 0)$ surveys and high-resolution followup from the MAGMA $^{12}$CO $(J = 1 \rightarrow 0)$ survey \citep{hug10} have mapped the molecular content of the entire disk of the LMC and identified 272 clouds that together contain $5 \times 10^7 \unit{M}_\odot$ of molecular gas.  When combined with multiwavelength archival observations of star formation indicators, these CO data constitute a snapshot in the evolution and star formation history of a population of GMCs.  

\citet{kaw09} used the NANTEN CO $J = (1 \rightarrow 0)$ data, along with complementary H$\alpha$ photometry \citep{kh86}, radio continuum maps at $1.4$, $4.8$, and $8.6 \unit{Ghz}$ \citep{dic05,hug07} and a map of young $(< 10\unit{Myr})$ clusters extracted from $UBV$ photometry \citep{bic96} to investigate the ongoing star formation within GMCs in the LMC.  These authors found a strong tendency for H~\textsc{ii} regions and young clusters to be spatially correlated with GMCs.  Using this association, the GMCs in their sample were separated into three types.  Type 1 GMCs are defined to be starless in the sense that they are not associated with detectable H~\textsc{ii} regions or young clusters, Type 2 GMCs are associated with H~\textsc{ii} regions, but not young clusters in the cluster catalog, and Type 3 GMCs are associated with both H~\textsc{ii} regions and young clusters.  $24 \%$ of the NANTEN sample were classified as Type 1, $50 \%$ as Type 2, and $26 \%$ as Type 3.  

Assuming that GMCs and clusters are formed in steady state and assuming that young clusters not associated with GMCs are associated with GMCs that have dissipated, one can infer from the NANTEN population statistics that GMCs spend $6 \unit{Myr}$ in the Type 1 phase, $13 \unit{Myr}$ in the Type 2 phase, $7 \unit{Myr}$ in the Type 3 phase, and then dissipate within $3\unit{Myr}$.  This accounting implies GMC lifetimes of approximately $20 {\rm to} 30\unit{Myr}$.  In support of the claim that the GMC classification scheme constitutes an evolutionary sequence, the authors note that among the resolved GMCs in the NANTEN survey, Type 3 GMCs are on average more massive, have larger turbulent line widths, and have larger radii.  However, there is significant scatter in the Type 3 GMC sample and the mass and size evolution are well within their error bars.  

In order to correct for extinction, which might obscure H$\alpha$ emitting H~\textsc{ii} regions, radio continuum maps at three, well-separated frequencies were used to identify obscured H~\textsc{ii} regions via their flat spectral slopes.  However, no H~\textsc{ii} regions were identified in the radio continuum data that were not present in the H$\alpha$ maps, leading the authors to conclude that the H$\alpha$ data was unaffected by obscuration.  No similar analysis was performed to estimate obscuration of young star clusters.  No attempt was made to correct for the varying sensitivities in the different radio maps, allowing for the possibility that some {\hii} regions were detected at 1.4 Ghz but below the sensitivity limit at 4.8 and 8.6 Ghz.

There are several observational biases inherent in the GMC classification scheme described above.  The first is the probable existence of star clusters and {\hii} regions located either behind or within giant molecular clouds from our viewpoint.  High dust extinction along these sightlines would mask some young clusters from detection in the \citet{bic96} star cluster sample.  This could lead to an overestimate of Type 2 GMCs relative to Type 3 GMCs.  Another possible bias is the use of the \citet{bic96} star cluster catalog.  Clusters in this catalog were targeted for $UBV$ photometry based on brightness and association with emission nebulae.  It is possible that some young clusters were missed in this catalog and no attempt is made by Kawamura et al. to correct for the completeness of the cluster catalog.  This would also lead to an overestimate of Type 2 GMCs relative to Type 3 GMCs. 

In order to make a quantitative comparison between our models and the evolutionary classification of \citet{kaw09}, we employ a few simple prescriptions to generate synthetic $V$ band, H$\alpha$, and radio continuum photometry for the clusters and H~\textsc{ii} regions in our simulated clouds. First, we calculate the $V$-band luminosity of our simulated clusters using the synthetic photometry of \citet{ls01}. For $5 \unit{M}_\odot \le M_* \le 15 \unit{M}_\odot$, the photometry is based on the evolution tracks of \citet{sch93}.  For massive stars, $M_* \ge 15 \unit{M}_\odot$, the synthetic photometry is based on the high mass-loss models of \citet{mey94}.  For both sets of synthetic photometry, we assume $Z = 0.008$.   Following \citet{par03}, we approximate variations in $L_V$ for these stars by only considering the main sequence evolution and taking $L_V = \left\langle L_V \right\rangle_{\rm MS}$, the mean luminosity on the main sequence.  Since the stellar evolutionary tracks give the luminosity at a discrete set of masses, we interpolate by fitting a broken power law between stellar masses with evolutionary tracks.  

We calculate the H$\alpha$ luminosity of our model {\hii} regions via \citep{mw97},
\begin{equation}
L_{\rm H\alpha} = 1.04 \times 10^{37} S_{49} \unit{erg}\unit{s}^{-1},
\label{halphalum}
\end{equation}
where $S_{49}$ is the ionizing luminosity of the central star cluster in units of $10^{49}\unit{photon}\unit{s}^{-1}$.  This is larger than the empirical relation by a factor of $1.37$ to correct for the absorption of ionizing radiation by dust grains.  Lastly, we find the radio continuum luminosity of our simulated H~\textsc{ii} regions via \citep{con92},
\begin{equation}
L_{\nu} = 1.6\times10^{23}S_{49}\left( \frac{1\unit{Ghz}}{\nu}\right)\unit{erg}\unit{s}^{-1}\unit{Hz}^{-1}
\label{radiolum}
\end{equation}
We also account for the reduction in flux from GMCs that would be larger than the beam size used by \citet{dic05} by performing a geometric correction and assuming that all {\hii} regions are placed at the center of the model beam.

\begin{figure}
\resizebox{\columnwidth}{!}{\includegraphics{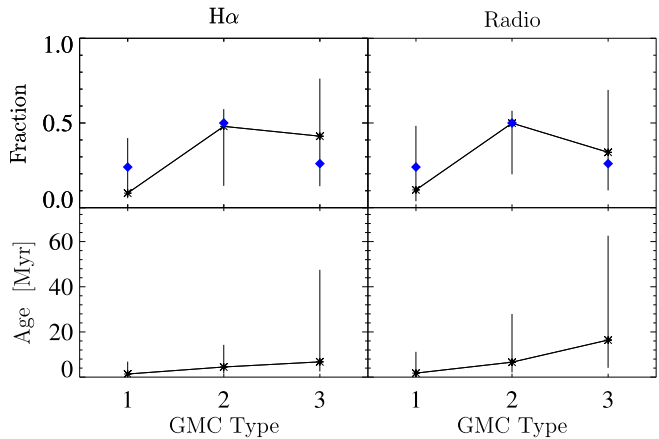}}
\caption{Fraction of GMC lifetime spent as a Type 1, 2, and 3 GMC as defined by \citet{kaw09} (\emph{diamonds, top row}), and average age of GMCs in each classification bin (\emph{bottom row}).  The asterisks indicate the median of the GMC type probability distribution functions generated using a Monte Carlo analysis described in the text.  The error bars encompass the 10th to 90th percentile interval of the probability distribution functions.}
\label{typecompare}
\end{figure}

To assign our simulated GMCs an evolutionary classification, we extract the ionizing luminosity and $V$ band luminosity of the brightest cluster in the GMC as a function of time.  Using the ionizing flux, we calculate the expected H$\alpha$ and radio continuum luminosity via equations~(\ref{halphalum}) and~(\ref{radiolum}) respectively.  For the $V$ band and $H\alpha$ luminosity, we correct for the foreground extinction, $A_V = 0.25$, towards the LMC \citep{sch98}, ignoring extinction internal to the LMC but external to the cloud under consideration.  To identify the GMC as either Type 2 or Type 3, we use either an H$\alpha$ luminosity cutoff or a radio continuum flux cutoff corresponding to the detection limits quoted by \citep{kaw09}.  For H$\alpha$, we say that an H~\textsc{ii} region is detected if $L_{H\alpha} \ge 10^{36}\unit{erg}\unit{s}^{-1}$.  For radio continuum, we say an H~\textsc{ii} region is detected if the radio flux at a distance of $50 \unit{kpc}$ would be greater than $0.7\unit{mJy}\unit{beam}^{-1}$ for a beam size of 20$''$  at $4.8\unit{GHz}$ \citep{dic05}.  Finally, we say that a GMC is Type 3 if it meets the criteria just described for H$\alpha$ or radio continuum as well as if $L_V \ge 1.66\times10^4 \unit{L}_\odot$, the completeness limit quoted by \citet{bic96} for the young cluster sample.  

Since there are bound to be clusters that are located both behind and within clouds along our line of sight, we also correct the synthetic $V$-band and H$\alpha$ photometry for extinction by the GMC.  This is done by assuming that star clusters form at random locations within clouds.  We calculate the optical depth to the star cluster via $\tau_\nu = \kappa_\nu \xi_\mu R_{\rm cl}$ where $\kappa_{\nu}$ is the mean dust opacity through the cloud, $\xi_\mu$ is the depth into the cloud in units of the cloud radius and $\mu$ is the angle between the normal to the surface of the cloud and the line of sight \citep{kmt08}.  For this purpose, we use a Milky Way extinction curve and assume an LMC dust-to-gas ratio whereby a column of $1 \unit{g} \unit{cm}^{-2}$ corresponds to $A_V = 107$.  In the visual passbands we are concerned with here, the Milky Way and LMC extinction curves are nearly identical.  

We have run a set of 2000 cloud models, 1000 each evolved using two different values of the ambient surface density, $\Sigma_{\rm res} = 8 \text{ and } 16 \unit{M}_\odot \unit{pc}^{-2}$.  All other parameters are as in Table~\ref{cloudparamtable}.  The resulting cloud models encompass the entire observed range in cloud masses reported by \citet{fuk08}.  

To directly compare to the observed distribution of GMC types we perform simulated observations using a Monte Carlo scheme.  Since the observations are inherently weighted by the GMC mass function, we first generate cloud masses by drawing from a powerlaw GMC mass spectrum with a slope of -1.6, a minimum mass of $5 \times 10^4 \unit{M}_{\odot}$ and a maximum mass of $5 \times 10^6 \unit{M}_\odot$ \citep{fk10}.  Once a mass is generated, we find all time steps where model clouds have masses within 0.1 dex of the randomly selected mass.  Within this sample of time steps, we calculate $f_1$, $f_2$, and $f_3$, the fraction of Type 1, Type 2, and Type 3 GMCs, respectively.  At the same time, we calculate $t_1$, $t_2$, and $t_3$ the average age of clouds in each GMC Type bin.  We generate $10^4$ Monte Carlo realizations, from which we construct probability distributions for $f_1$, $f_2$, $f_3$, $t_1$, $t_2$, and $t_3$.   

The results of this comparison are presented in the top row of Figure~\ref{typecompare}.  In the figure, the lines connect the median of the Monte Carlo probability distributions while the error bars encompass the 10th and 90th percentile.  We are able to reproduce the observed distribution of Type 1, 2 and 3 GMCs as observed by \citet{kaw09}.  In particular, using both detection limits, we find that the majority of clouds are detected as Type 2 GMCs, and relatively fewer clouds are detected as Type 1 and 3 GMCs.  Interestingly, in the bottom panel of the figure, we find that, on average, the GMC classification scheme does constitute an age sequence in that Type 2 GMCs tend to be somewhat older than Type 1 clouds.  Type 3 GMCs in turn tend to be older than Type 2 clouds.  On the other hand, the spread in cloud ages within each bin is well within the error bars, indicating that the GMC type classification is not necessarily a strict evolutionary sequence: older clouds can be classified as Type 1 and younger clouds can be classified as Type 3.

\begin{figure}
\resizebox{\columnwidth}{!}{\includegraphics{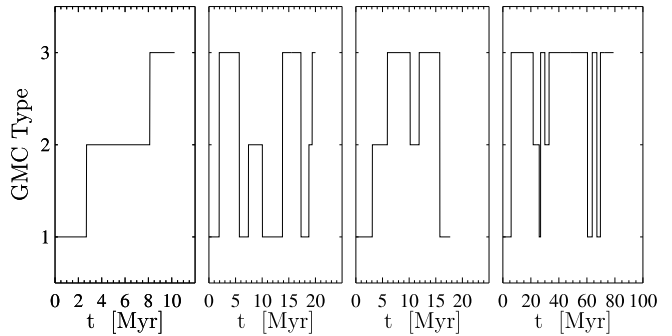}}
\caption{GMC classification as a function of time for a selection of model clouds.}
\label{typehistory}
\end{figure}

This can be seen more clearly in Figure~\ref{typehistory}, where we plot the classification of a selection of GMCs as a function of time.  We see that for some clouds, the classification scheme does represent an evolutionary sequence.  In these runs, thw cloud starts as a Type 1 GMC, begins forming star clusters, evolves into a Type 2 GMC, and then forms massive OB association and becomes a Type 3 GMC.  However, we also see that a cloud can quickly form a massive OB association and be classified as a Type 3 GMC early on and only later be classified as a Type 2 GMC.  Alternatively, a cloud may happen to not form any massive clusters late in its evolution, causing a massive and old cloud to be identified as a Type 2 or 1 GMC late in its evolution.  Finally, there are clouds which exhibit no discernible pattern in their histories, more or less randomly transitioning between GMC classifications throughout their lives.  This may explain the presence of massive ($\sim 10^6 \unit{M}_\odot$) Type 1 ``young" GMCs in the samples of \citet{kaw09} and \citet{hug10}.

\section{Caveats and Limitations}
\label{caveats}

\subsection{Implications of the Assumption of Homology}

Assuming that clouds evolve homologously is the main limitation of our model.  We make the assumption of homology to significantly simplify the equations governing the evolution of the cloud.  Given the assumption of homology, our equations of motion follow rigorously from the local equation of momentum and energy conservation.  A more complex cloud structure destroys the relative simplicity of the model and would require computationally expensive hydrodynamical simulations to model accretion in detail.  

Homology constrains the cloud to always maintain the same shape and degree of central concentration.  This is equivalent to setting time derivates of the structure constant, $\aone$, to zero.  This might be a problem if changes of the moment of inertia of clouds occur primarily by changing the shape of the cloud rather than through overall expansion or contraction.  It might also be a problem if the accretion flow does not in reality differentially mix with the cloud.  If the accretion flow is anisotropic or if it cannot fall to the central regions of the cloud before it mixes, the cloud may become less centrally condensed and we may overestimate the kinetic energy injection since material cannot fall to the bottom of the cloud potential well.

While we cannot resolve the dynamical effect of changes in the shape of the cloud, we can resolve changes in the size of the cloud.  Given the observed Larson scaling relations, we expect more massive clouds to be larger, implying that clouds must expand as they accrete mass.  If clouds do form via gravitational instability, they must accrete significant mass.  We do resolve this behavior in our models.  We caution that our virial analysis most readily describes a relaxed system.  We may be doing a poor job of resolving phenomena that occur on a timescale comparable to the crossing time.

\subsection{Magnetic fields in the Atomic Envelope}

One key assumption we made in deriving the global energy equation was that the atomic envelope contributes negligibly to the total magnetic energy associated with the cloud.  That is, we did not include the magnetic field when calculating $\mathcal{E}_{\rm amb}$ in equation~(\ref{eambdef}).  The motivation for this assumption is based on comparisons of observations of magnetic fields in dense molecular clumps, where it is possible to measure the magnetic field directly via the Zeeman effect in lines of OH and CN \citep{tro08,fal08}, to the magnetic fields in the atomic ISM, measured via the Zeeman effect in the 21 cm line of neutral Hydrogen \citep{hei05}.  These studies consistently find that the magnetic field strength is significantly elevated in the dense molecular gas.  However, most of the volume of a GMC is occupied by diffuse molecular gas with low OH abundance --- frustrating efforts to measure magnetic fields via observations of OH in emission.  Thus, without direct observations of magnetic fields in the diffuse molecular gas, we cannot know whether the magnetic fields in the atomic envelope are weak or strong compared to the magnetic field strength in the bulk of a GMC.  While preliminary observations by Troland et al. (2011, private communication) indicate that magnetic fields in the diffuse molecular component are somewhat stronger than in the atomic envelope, this question has yet to be settled.  It is possible that we underestimate the net magnetic energy due to the presence of the cloud.

Another possible problem with our treatment of magnetic fields is that we assume the mass-to-flux ratio remains constant throughout the evolution of the cloud.  This is equivalent to assuming a fixed value for $\eta_B$.  This assumption might be invalid if accreted material flows preferentially along magnetic field lines or if ambipolar diffusion can act act on timescales comparable to the cloud dynamical time.  While measurements \citep{li061} and theoretical esimates \citep{mck07} of the mass-to-flux ratio of molecular clouds find that clouds should be marginally supercritical, with mass-to-flux ratios of order unity, the time-evolution and cloud-to-cloud variation in mass-to-flux ratios are poorly constrained.

\subsection{Validity of Larson's Laws}

It has been suggested that the observed constancy of GMC surface densities is an artifact \citep[see e.g.][]{keg89, vaz97, bal02, bal06,bal11}.  Since the physical structure we assume for the clouds in our model implicitly assumes that the Larson relations are valid, this may imply that our models do not correspond well to real GMCs.  

The argument that the Larson laws are a product of a selection effect usually proceeds as follows.  If one looks for overdensities of a particular size in a simulation of the turbulent ISM, one will find that the selected clouds have a wide distribution of masses, implying a large spread in cloud-to-cloud surface density.  Since observers detect clouds using CO as a tracer and at low surface density the CO abundance is not high enough to be detectable in emission, observers will never find low surface density clouds.  This implies that at a fixed radius, real clouds should have more variation in mass than a naive interpretation of the CO observations would suggest.

This argument misses two key aspects of the observed properties of GMCs.  The first is that it cannot explain the lack of clouds at {\it high} gas surface densities.  For the same reason that we should not be able to see diffuse clouds, we should very easily be able to see compact, high surface density CO clouds.  The fact that these clouds don't exist implies something important about molecular cloud structure.  The second argument is that the lack of low surface density clouds does not imply that molecular clouds can exist at all surface densities but merely that the molecular clouds dissociate once they become optically thin to the ambient ultraviolet radiation field.  While diffuse atomic clouds certainly exist, these clouds do not form stars \citep{klm11}.

Since high surface density GMCs are not observed in the local universe and low surface density clouds are not molecular and thus not GMCs by definition, the observed lack of variation in GMC surface densities must be a real property of the clouds.  This has been confirmed with very detailed dust extinction measurements of nearby star forming clouds, where an exquisitely tight mass-radius relation is observed \citep{lom10} and in extragalactic studies where little variation is seen when comparing the mass-radius relation from galaxy to galaxy \citep{bol08}.  Taken together, the evidence seems to imply that the Larson relations are a property of the structure of GMCs and are not due to a selection effect.

\section{conclusions}

In this paper, we have presented semianalytic dynamical models for the evolution of giant molecular clouds undergoing both mass accretion and star formation.  These models are able to capture the evolution of individual GMCs from their growth and the onset of massive star formation, until their dispersal via an energetic H~\textsc{ii} region or through the combined action of accretion and star formation.  We are able for the first time to synthesize galactic populations of GMCs whose properties correspond closely to the observed properties of GMCs in the Milky Way and nearby external galaxies.  We have shown that clouds in low surface density environments generally disperse within a few crossing times, before they can accrete all of the gas in their reservoir.  At the same time, clouds in high surface density environments do accrete all of the gas in their reservoirs and tend to be larger and more massive.   We have also shown that mass accretion can contribute a significant fraction of the total energy available for turbulent driving.  Lastly, we generate synthetic cluster observations and compare against the evolutionary classification scheme of \citet{kaw09}, finding good agreement when we correct for selection effects and systematic biases inherent in the observations.  We conclude that, on average, the evolutionary classification scheme corresponds to an age sequence but is not a good predictor for the evolutionary state of isolated clouds.

\acknowledgements

We would like to thank Leo Blitz and Adam Leroy for making available the Milky Way data used in Figures~\ref{blitzplot} and~\ref{blitzplot_lw}.  We would also like the thank an anonymous referee for their helpful comments which spurred a significant improvement in the quality of the paper. NJG is supported by a Graduate Research Fellowship from the National Science Foundation.  MRK acknowledges support from: an Alfred P. Sloan Fellowship; NSF grants AST-0807739 and CAREER-0955300; and NASA through Astrophysics Theory and Fundamental Physics grant NNX09AK31G and a Spitzer Space Telescope Theoretical Research Program grant. CDM's research is supported by an NSERC discovery grant and an Ontario Early Researcher Award. The research of CFM is supported in part by NSF grant AST-0908553 and by a Spitzer Space Telescope Theoretical Research Program grant.

\begin{appendix}
\section{Derivation of the Virial Theorem for an accreting cloud}
\label{derivationEVT}
Here we derive an equation governing the virial balance of a cloud that is simultaneously forming stars and accreting material.  This is a generalization of the analysis of Paper I and \citet{mz92}.  We refer the reader to those papers for details that are unrelated to the accretion flow and to \S\,\ref{overview} for a general overview of the model.

Consider a single molecular cloud contained within an Eulerian volume $V_{\mathrm{vir}}$ with bounding surface $S_{\mathrm{vir}}$.  We assume that $V_{\mathrm{vir}}$ is sufficiently large to contain the cloud at all times.  Material within the virial volume is apportioned into three components: virial material, a gaseous reservoir, and material in a photoionized wind.  Locally, each component satisfies its own continuity equation,

\begin{align}
\frac{\partial \rho}{\partial t} &= - \div{ \rho \mathbf{v}} + \dot{\rho}\\
\frac{\partial \rho_{\rm res}}{\partial t} &= -\div{\rho_{\rm res} \mathbf{v_{\rm res}}} - \dot{\rho}_{\rm acc}\\
\frac{\partial \rho_{\rm w}}{\partial t} &= -\div{ \rho_{\rm w} \mathbf{v_{\rm w}}} - \dot{\rho}_{\rm ej}.
\end{align}

\noindent
These equations are coupled via the source and sink terms,

\begin{equation}
\dot{\rho} = \dot{\rho}_{\rm acc} + \dot{\rho}_{\rm ej}.
\label{rhodotdef}
\end{equation}

\noindent
We assume that accretion can only transport material onto the cloud and that the wind can only carry material away from the cloud, implying $\dot{\rho}_{\rm acc} \ge 0$ and $\dot{\rho}_{\rm ej} \le 0$.

The local equation of momentum conservation for the virial material is (c.f. equation [A2] in Paper I)
\begin{equation}
\frac{\partial}{\partial t}(\rho \mathbf{v}) = -\div{ (\Pi - \mathbf{T}_{\rm M})} + \rho\mathbf{g} + \mathbf{F}_* + \dot{\rho}_{\rm ej}(\mathbf{v} + \mathbf{v}_{\rm ej}^\prime) + \dot{\rho}_{\rm acc}\mathbf{v}_{\rm res}
\label{momeq}
\end{equation}
where $\Pi = P_{\rm th}\mathbf{I} + \rho\mathbf{v}\mathbf{v} - \mathbf{\pi}$ is the gas pressure tensor, $\pi$ is the viscous stress tensor, $\mathbf{T}_{\rm M} = [\mathbf{B}\mathbf{B} - (1/2)B^2\mathbf{I}]/(4\pi)$ is the Maxwell stress tensor, $\mathbf{B}$ is the magnetic field, $\mathbf{I}$ is the unit tensor, $\mathbf{g}$ is the gravitational force per unit mass, and $\mathbf{F}_* + \dot{\rho}_{\rm ej}(\mathbf{v}+\mathbf{v}_{\rm ej}^\prime)$ is the local body force due to the interaction between expanding H~\textsc{ii} regions and the cloud.  We write the stellar forcing term in this form so as to separate the random component, $\mathbf{F}_*$, from the spherically symmetric component $\dot{\rho}_{\rm ej}(\mathbf{v} + \mathbf{v}_{\rm ej}^\prime)$.  

$\mathbf{F}_*$ is a function of time and position in the cloud and depends on the precise location and history of massive star formation.  Due to the nature of supersonic turbulence, we expect turbulent overdensities to be randomly scattered throughout the cloud.  Thus, we expect $\mathbf{F}_*$ to be randomly oriented with respect to the position vector, implying $\int_{V_{\rm vir}} \mathbf{r}\cdot \mathbf{F}_* dV = 0$.  While $\mathbf{F}_*$ is randomly oriented, $\dot{\rho}_{\rm ej}(\mathbf{v}+\mathbf{v}_{\rm ej}^\prime)$ should on average be purely radial because the photoionized wind will be blown out preferentially along the pressure gradient.  Neither $\fstar$ nor the viscous stress tensor is included in the momentum equation used in Paper I.  A detailed comparison of the two approaches leads us to conclude that the formulation used here more properly follows the transport of energy through the turbulent cascade.

After taking the second time derivative of the cloud moment of inertia, $I_{\rm cl} = \int_{\vvir}\rho r^2 dV$, and substituting the momentum equation~(\ref{momeq}) into the resulting expression, we find
\begin{equation}
\begin{split}
\frac{1}{2}\ddot{I}_{\rm cl} &= \frac{1}{2}\frac{d}{dt}\int_{\vvir} \rhodot r^2 dV - \frac{1}{2}\frac{d}{dt}\int_{\svir} \rho \mathbf{v} r^2 \cdot d \mathbf{S} +  \int_{\vvir} \mathbf{r} \cdot [\rho \mathbf{g} + \fstar - \div{(\Pi - \mathbf{T}_{\rm M})}] dV \\
& + \int_{\vvir} \rhodotej \mathbf{r} \cdot (\mathbf{v} + \vej)dV + \int_{\vvir} \rhodotacc \mathbf{r} \cdot \vres dV.
\label{virialstep}
\end{split}
\end{equation}

\noindent
Upon evaluating the integrals in equation~(\ref{virialstep}) term by term, we obtain the Eulerian Virial Theorem,
\begin{equation}
\begin{split}
\label{EVT}
\frac{1}{2}\ddot{I}_{\rm cl} &= 2(\mathcal{T} - \mathcal{T}_0) + \mathcal{B} + \mathcal{W} - \frac{1}{2}\frac{d}{dt}\int_{S_{\rm vir}} (\rho \mathbf{v} r^2) \cdot d\mathbf{S} + a_\textrm{I}\dot{M}_{\rm cl}R_{\rm cl}\dot{R}_{\rm cl}\\
 &+ \frac{1}{2}a_\textrm{I}\ddot{M}_{\rm cl}R_{\rm cl}^2 + a_\textrm{I}\dot{M}_{\rm ej}R_{\rm cl}\dot{R}_{\rm cl} + \frac{3-k_\rho}{4-k_\rho}R_{\rm cl}(\dot{M}_{\rm ej}v_{\rm ej}^\prime - \xi\dot{M}_{\rm acc}v_{\rm esc})
\end{split}
\end{equation}

\noindent
Here $\mathcal{T}$, $\mathcal{T}_0$, $\mathcal{B}$, and $\mathcal{W}$ are respectively the standard kinetic, surface kinetic, magnetic, and gravitational terms (see Paper I for precise definitions) and $\aone = (3-\krho)/(5-\krho)$.  The final term in equation~(\ref{EVT}) does not appear in the EVT derived in Paper I and is due to the presence of the accretion flow.  This has the same form as the wind recoil term except for the presence of a dimensionless factor

\begin{equation}
\xi = \int_0^1 (4-\krho)x^{3-\krho}\left\{1 + f \left[\frac{1-x^{2-\krho}}{2-\krho} + \int_x^1 \frac{y(x^\prime)}{x^{\prime 2}}dx^\prime \right] \right\} dx
\label{xidef}
\end{equation}

\noindent
which arises because material is accreted at a velocity that depends on the depth of the cloud potential well.  The dimensionless variables $x, y, \text{ and } f$ are defined explicitly in Appendix~\ref{accretionappendix}.

The magnetic term $\mathcal{B}$ retains the form used in Paper I and derived by \citep{mz92} because we assume any deformation of the magnetic field in the ambient medium caused by the presence of the cloud is negligible at the virial surface, allowing us to approximate that the virial volume is threaded by a constant magnetic field $B_0$. Here $B_0$ is the RMS value of the true magnetic field at the virial surface, which may fluctuate around $B_0$.  Since material in the reservoir should also carry currents and thus generate magnetic fields, this parameterization underestimates the total magnetic energy.  However, we expect the mean density of reservoir material within the virial volume to be an order of magnitude smaller than the density of cloud material (see Appendix~\ref{accretionappendix}), so the contribution of the reservoir to the magnetic energy is small compared to the contribution due to the cloud.

\section{Properties of the Reservoir}
\label{accretionappendix}
Consider an accreting cloud that is not forming stars and thus not generating a wind.  Let $M_{\rm res}(r,t)$ be the mass of material in the accretion flow contained within a radius $r$ at time $t$ and let $\Delta r = v_{\rm res,sys}\Delta t$ be the distance that the accreting gas falls in a time $\Delta t$.  In the same time, a fraction of the material in the accretion flow will be converted into cloud material. In the frame comoving with the accretion flow, the change in the mass of reservoir interior to a radius $r$ in a time $\Delta t$ is,

\begin{equation}
M_{\rm res}(r,t) - M_{\rm res}(r - \Delta r,t + \Delta t) = -\Delta t \int_0^r \dot{\rho}_{\rm acc} dV.
\label{mresdef}
\end{equation}

\noindent
Upon Taylor expanding in $\Delta t$ and $\Delta r$, dropping the nonlinear terms, and evaluating the integral on the right hand side of equation~(\ref{mresdef}), we find,

\begin{equation}
\Delta r \frac{\partial M_{\rm res}(r,t)}{\partial r} - \Delta t \frac{\partial M_{\rm res}(r,t)}{\partial t} = -\dot{M}_{\rm acc}(r,t) \Delta t.
\end{equation}

\noindent
If the accretion flow is in quasi-steady state, the time derivative vanishes and $M_{\rm res}(r,t) = M_{\rm res}(r)$.  Integrating to obtain $M_{\rm res}(r)$, we find

\begin{equation}
M_{\rm res}(r) = - \int_0^r \frac{\dot{M}_{\rm acc}(r^\prime)}{v_{\rm res,sys}(r^\prime)} dr^\prime
\label{mresdef}
\end{equation}

If we wish to evaluate the above integral and obtain an expression for $M_{\rm res}(r)$, we need to know $v_{\rm res,sys}(r)$.  Expanding equation~(\ref{vressysdef}), we see

\begin{equation}
\frac{1}{2}v_{\rm res,sys}^2(r) = -\int_\infty^{\rcl} \frac{G(\mcl + \mres)}{r^{\prime 2}} dr^\prime - \int_{\rcl}^r \frac{G \mcl(r^\prime)}{r^{\prime 2}}\left[\left(\frac{r^\prime}{\rcl}\right)^{3-\krho} + \frac{\mres(r^\prime)}{\mcl(r^\prime)} \right]dr^\prime.
\label{vindef}
\end{equation}

\noindent
Equations~(\ref{mresdef}) and~(\ref{vindef}) constitute a system of integral equations that must be simultaneously solved to obtain a solution for the velocity and mass profile of reservoir material within the cloud volume.  If we define the functions $x = r/R_{\rm cl}$, $y(x) = M_{\rm res}(R_{\rm cl}x)/M_{\rm cl}$, and $f = M_{\rm cl}/(M_{\rm res}+M_{\rm cl})$, equation~(\ref{vindef}) reduces to
\begin{equation}
v_{\rm res,sys}(r) = -v_{\rm esc}\left[1+f\left(\frac{1 - x^{2-k_{\rho}}}{2-k_{\rho}} - \int_{1}^{x}\frac{y(x^{\prime})}{x^{^{\prime}2}}dx^{\prime}\right)\right]^{1/2}.
\label{vineqn}
\end{equation} 

\noindent
Defining $\zeta = \dot{M}_{\rm acc}t_{\rm ff}/M_{\rm cl}$ and $z(x) = \int_{x}^{1}y(x^\prime)/x^{\prime 2}dx^{\prime}$, we see that the problem of determining both $v_{\rm res,sys}(r)$ and $M_{\rm res}(r)$ reduces to solving the following system of nonlinear ordinary differential equations:

\begin{align}
\frac{dy}{dx} &= \frac{2}{\pi}\zeta x^{3-k_\rho}\left[1+f\left(\frac{1-x^{2-k_\rho}}{2-k_\rho}+z(x)\right)\right]^{-1/2},\label{reservoirequation1}\\
\frac{dz}{dx} &= -\frac{y(x)}{x^2}.
\label{reservoirequation2}
\end{align}

These equations can be solved numerically using the shooting method as follows.  Since there should be no unmixed accreted material at $r=0$, we expect $y(0) = 0$.  Since $z(x)$ is defined in terms of an integral, we have no {\it a priori} knowledge of $z(0)$.  However, we do know that $z(1) = 0$ and we expect $y(1) \le \zeta$.  If we impose $y(1) = c\, \zeta$ where $c$ is a constant of order unity, we can integrate the above system and obtain a trial solution for $y(x)$ and $z(x)$.  The constant $c$ can then be varied until a solution is obtained with $y(0) = 0$.  There is a unique solution for each value of $\zeta$ and thus a new solution must be obtained if $\zeta$ varies.   

\begin{figure}
\resizebox{\textwidth}{!}{\includegraphics{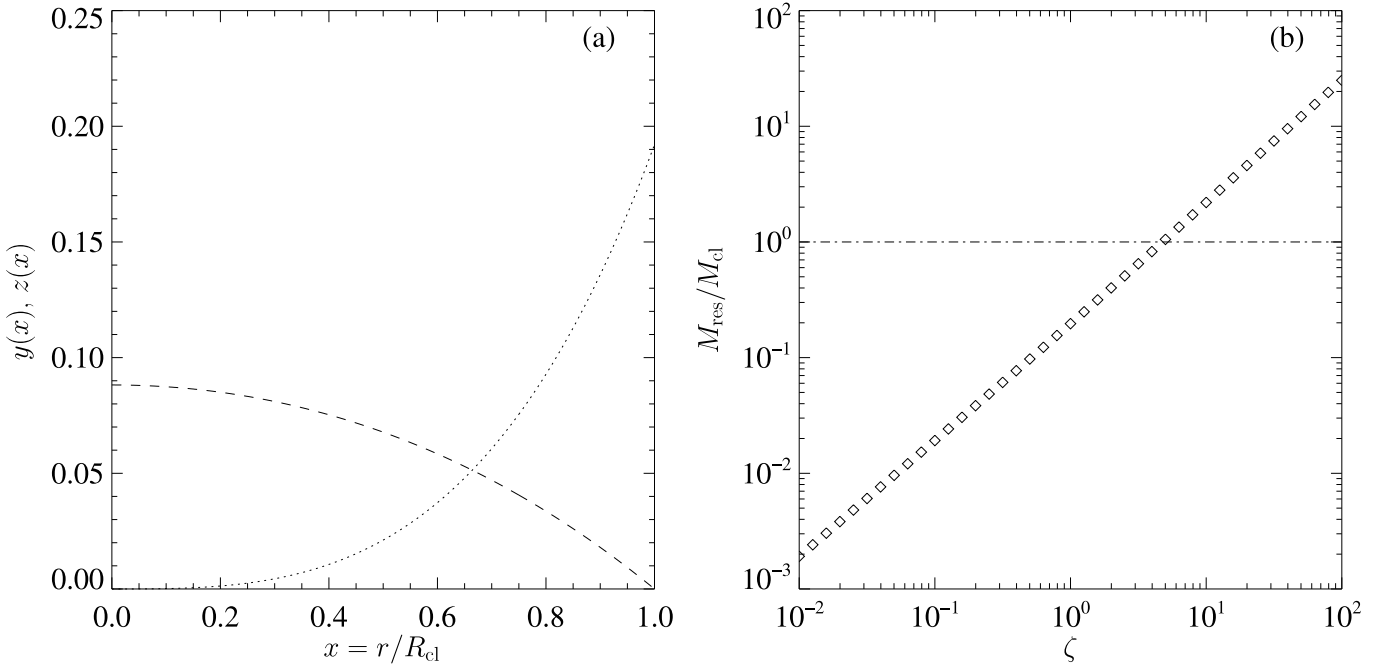}}
\caption{(a): $y$ (dotted line) and $z$ (dashed line) as a function of $x = r/R_{\rm cl}$.  (b): $y(1) = M_{\rm res}(R_{\rm cl}) / M_{\rm cl}$ as a function of $\zeta = \dot{M}_{\rm acc}t_{\rm ff} / M_{\rm cl}$.  We see $M_{\rm res}(R_{\rm cl}) \ge M_{\rm cl}$ for $\zeta \gtrsim 5$.}
\label{appendix2plot}
\end{figure}

A key assumption in this analysis was that the accretion rate is approximately constant over the cloud free-fall timescale.  This is equivalent to the assumption that $\zeta \lesssim 1$.  This is a reasonable assumption, which we can see by making an analogy to the case of a protostellar core.  Since we expect $\dot{M}_{\rm cl} \approx M_{\rm cl}/t_{\rm ff,r}$ where $t_{\rm ff,r}$ is the free-fall time for all of the gas in the reservoir \citep{sta80,mt03} and since the reservoir should be significantly more extended than the cloud, we expect the mean density of the reservoir to be much lower than the mean density in the cloud.  This implies $t_{\rm ff,r} \gg t_{\rm ff,cl}$.  If this condition does hold, then the condition $\zeta \le 1$ is automatically satisfied.

In Figure~\ref{appendix2plot}a we present a numerical solution for $y(x)$ and $z(x)$ obtained for $\zeta = 1.0$.  We see for this case that $M_{\rm res}(R_{\rm cl}) \approx 0.2 M_{\rm cl}$, showing that even for substantial accretion rates, the gas within $\vcl$ is primarily composed of cloud material.  In Figure~\ref{appendix2plot}b we show $y(1) = M_{\rm res}(R_{\rm cl}) / M_{\rm cl}$ as a function of $\zeta$.  Reservoir material becomes the primary component of the volume occupied by the cloud for $\zeta \gtrsim 5.0$.  For this reason, we reject as unphysical any portion of cloud history with $\zeta \ge 5$.  In practice, we find $\zeta \lesssim 1.0$ for the lifetime of all clouds we simulate.

\section{Derivation of the Equation of Energy Conservation for an Accreting Cloud}
\label{derivationEnergy}
Here we derive the global equation of energy conservation for a cloud undergoing accretion and star formation.   We begin by writing the equation of momentum conservation (equation~[\ref{momeq}]) in Lagrangian form,

\begin{equation}
\rho \frac{d\mathbf{v}}{dt} = -\grad{ P} - \rho \grad{ \phi } + \div{ \pi} + \frac{\mathbf{J} \times \mathbf{B}}{c} + \dot{\rho}_{\rm ej}\mathbf{v}_{\rm ej}^\prime +\mathbf{F}_* + \dot{\rho}_{\rm acc}(\mathbf{v}_{\rm res} - \mathbf{v}).
\end{equation}

\noindent
Upon contracting the above equation with $\mathbf{v}$, we obtain the nonthermal energy evolution equation,

\begin{equation}
\begin{split}
&\frac{\partial}{\partial t}\left(\frac{1}{2}\rho v^2 + \rho \phi_{\rm cl} + \frac{B^2 - B_0^2}{8\pi}\right) + \div{ \rho \mathbf{v}\left(\frac{1}{2}v^2 + \phi_{\rm cl}\right)} +  \div{ \mathbf{S}_{\rm P}} = - \mathbf{v}\cdot \grad{ P} + \rho \frac{\partial \phi_{\rm cl}}{\partial t} + \rho\mathbf{v}\cdot\mathbf{g}_{\rm res} + \div{(\pi \cdot \mathbf{v})} \\ 
&- \pi:\grad{\mathbf{v}} + \dot{\rho}\left(\frac{1}{2}v^2+\phi_{\rm cl}\right) + \mathbf{v}\cdot \mathbf{F}_* + \dot{\rho}_{\rm ej} \mathbf{v} \cdot \mathbf{v}_{\rm ej}^\prime + \dot{\rho}_{\rm acc}(\mathbf{v}\cdot\mathbf{v}_{\rm res} - v^2)
\label{nonthermevol}
\end{split}
\end{equation}

\noindent
where $\mathbf{S}_{\rm P}$ is the Poynting vector.  Combining the first law of thermodynamics with the continuity equation yields the evolution equation for the thermal energy of the cloud,

\begin{equation}
\frac{\partial}{\partial t}(\rho e) + \div{\rho \mathbf{v} \left(e + \frac{P}{\rho}\right)} = \dot{\rho} e + \mathbf{v} \cdot \grad{ P} + \Gamma - \Lambda
\label{thermevol}
\end{equation}

\noindent
where $e$ is the internal energy per unit mass and $\Gamma$ and $\Lambda$ are respectively the rates of energy gain and loss per unit volume.  Here we've assumed that accreted material has the same thermal energy density as cloud material, which will be true if the radiative timescales are short compared to the mechanical timescales, as in molecular cloud conditions.  Summing equation~(\ref{nonthermevol}) and~(\ref{thermevol}), we obtain the evolution equation for the total energy of the system

\begin{equation}
\begin{split}
&\frac{\partial}{\partial t}\left[\rho\left( \frac{1}{2}v^2 + e + \frac{1}{2}\phi_{\rm cl}\right) + \frac{B^2 - B_0^2}{8\pi}\right] + \div{ \rho \mathbf{v}\left(\frac{1}{2}v^2 + e + \frac{P}{\rho} + \phi_{\rm cl}\right)} + \div{\mathbf{S}_{\rm P}} =  \frac{1}{2}\left(\rho \frac{\partial \phi_{\rm cl}}{\partial t} - \frac{\partial \rho}{\partial t}\phi_{\rm cl}\right) \\
& + \dot{\rho}\left(\frac{1}{2}v^2 + e + \frac{1}{2}\phi_{\rm cl}\right)  + \frac{1}{2}\rhodot\phi_{\rm cl} + \rho\mathbf{v} \cdot \mathbf{g}_{\rm res} + \dot{\rho}_{\rm ej} \mathbf{v} \cdot \mathbf{v}_{\rm ej}^\prime  +  \dot{\rho}_{\rm acc}(\mathbf{v}\cdot\mathbf{v}_{\rm res} - v^2) + \div{(\pi \cdot \mathbf{v})} + \mathbf{v}\cdot \mathbf{F}_*  - \Lambda.
\label{energyevol}
\end{split}
\end{equation}

\noindent
Here we've assumed that $\Gamma = \pi : \grad{ \mathbf{v}}$, the scalar rate of viscous dissipation.  This is equivalent to the statement that turbulent kinetic energy is converted into heat at viscous scales whereupon the energy is quickly radiated away.  While there may be other heating mechanisms, including cosmic rays and protostellar radiation, we neglect these sources by noting that any local heating should be offset in a time much shorter than the dynamical time.

If we define the total energy due to the presence of the cloud,

\begin{equation}
\mathcal{E_{\rm cl}} = \int_{V_{\rm cl}}\rho\left(\frac{1}{2}v^2 + e + \frac{1}{2}\phi_{\rm cl} \right)dV + \int_{V_{\rm vir}}\frac{B^2-B_0^2}{8\pi}dV
\label{ecldef}
\end{equation}

\noindent
and the total energy in the ambient medium,

\begin{equation}
\mathcal{E}_{\rm amb} = \int_{V_{\rm vir} - V_{\rm cl}} \rho \left( \frac{1}{2}v^2 + e + \frac{1}{2}\phi_{\rm cl} \right) dV,
\label{eambdef}
\end{equation}

\noindent
we can integrate equation~(\ref{energyevol}) over the virial volume and obtain an evolution equation for $\mathcal{E}_{\rm cl}$.  Portions of this calculation are performed explicitly in Paper I and we do not reproduce those results here.  The new terms stem from our inclusion of accretion as well as the inclusion of $\fstar$, the random component of the stellar forcing term.  The integrals over the new terms are evaluated below.

The first new term is due to the gravitational influence of the reservoir,
 
\begin{equation}
\int_{V_{\rm vir}}\rho\mathbf{v}\cdot\mathbf{g}_{\rm res}dV = -\chi\frac{\dot{R}_{\rm cl}}{R_{\rm cl}}\frac{GM_{\rm cl}^{2}}{R_{\rm cl}}
\end{equation}

\noindent
where

\begin{equation}
\chi = (3-k_\rho)\int_0^1 x^{1 - k_\rho} y(x) dx.
\label{chidef}
\end{equation}

\noindent
This is the gravitational work done on the cloud by the reservoir material as the cloud expands and contracts.  

The random stellar forcing term can be evaluated by noting that $\mathbf{r}$ and $\mathbf{F}_*$ are assumed to be uncorrelated, so $\int_{V_{\rm vir}} \mathbf{v} \cdot \mathbf{F}_*dV = \int_{V_{\rm vir}} \mathbf{v}_{\rm turb} \cdot \mathbf{F}_* dV$.  This integral depends on the degree of correlation between two randomly oriented vectors, $\mathbf{v}_{\rm turb}$ and $\mathbf{F}_*$.  Since, at a fixed time, the direction of expansion of an H~\textsc{ii} region is the same as the direction of momentum injection, these vectors should indeed be highly correlated.  This term then corresponds to the net injection of energy by H~\textsc{ii} regions,

\begin{equation}
\mathcal{G}_{\rm cl} = \int_{V_{\rm vir}} \mathbf{v}_{\rm turb} \cdot \mathbf{F}_* dV.
\label{feedbackterm}
\end{equation} 

For the terms proportional to $\dot{\rho}_{\rm acc}$, we have

\begin{equation}
\int_{V_{\rm vir}} \dot{\rho}_{\rm acc} \mathbf{v} \cdot \mathbf{v}_{\rm res}dV = -\left(\frac{3-k_{\rho}}{4-k_{\rho}}\right)\xi \dot{R}_{\rm cl}\dot{M}_{\rm acc}v_{\rm esc} + \int_{V_{\rm vir}}\dot{\rho}_{\rm acc}\mathbf{v}_{\rm turb}\cdot\mathbf{v}_{\rm res}dV
\label{accterm}
\end{equation}

\noindent
and

\begin{equation}
\int_{V_{\rm vir}}\dot{\rho}_{\rm acc}v^2dV =  a_{\textrm{I}}\dot{M}_{\rm acc}\dot{R}_{cl}^{2} + 3\dot{M}_{\rm acc}\sigma_{\rm cl}^2 
\end{equation}

\noindent
where we've used our assumption that $\mathbf{v}_{\rm res,rand}$ is randomly oriented with respect to $\mathbf{r}$.

We are left with one last integral on the right hand side of equation~(\ref{accterm}) that depends on the correlation between two vector fields, $\mathbf{v}_{\rm turb}$ and $\mathbf{v}_{\rm res}$. Suppose $\mathbf{v}_{\rm res}$ and $\mathbf{v}_{\rm turb}$ are perfectly correlated.  In that case, the integral we are concerned with can be readily evaluated in three limits, $|\vres| = |\vturb|$, $|\vres| \gg |\vturb|$, and $|\vres| \ll |\vturb|$.  In the first limit $|\vres| = |\vturb|$ the transfer of material to the cloud is merely an act of relabeling, so $\mathbf{v}_{\rm turb}\cdot\mathbf{v}_{\rm res} = \mathbf{v}_{\rm res}^2$.  Now consider the limit $|\vres| \gg |\vturb|$.  We approximate that a parcel of reservoir material mixes with the cloud once it has swept up a mass of cloud material equal to its own mass.  Since the interaction between cloud material and reservoir material must be inelastic if they are to mix, the velocity of cloud material must be driven by the act of mixing to $\vturb = \vres/2$.   Conversely, if $|\vres| \ll |\vturb|$, the reservoir material is driven to a velocity $\vres = \vturb/2$.  We obtain the correct answer in all three limits if we assume
\begin{equation}
\int_{V_{\rm vir}}\dot{\rho}_{acc}\mathbf{v}_{\rm turb}\cdot\mathbf{v}_{\rm res}dV = \frac{1}{2}\int_{V_{\rm vir}}\dot{\rho}_{acc}(\mathbf{v}_{\rm res}^2 + \mathbf{v}_{\rm turb}^2)dV. 
\end{equation}
This is an upper bound on the value of the integral in the limit of perfect correlation between $\mathbf{v}_{\rm turb}$ and $\mathbf{v}_{\rm res}$.  If $\mathbf{v}_{\rm turb}$ and $\mathbf{v}_{\rm res}$ are uncorrelated then the integrand should on average be zero.  Thus we have,
\begin{equation}
0 \le \int_{V_{\rm vir}}\dot{\rho}_{acc}\mathbf{v}_{\rm turb}\cdot\mathbf{v}_{\rm res}dV \le \frac{1}{2}\int_{V_{\rm vir}}\dot{\rho}_{acc}(\mathbf{v}_{\rm res}^2 + \mathbf{v}_{\rm turb}^2)dV.
\end{equation}
To evaluate this integral, we linearly interpolate between the upper limit and lower limit,
\begin{equation}
\int_{V_{\rm vir}}\dot{\rho}_{acc}\mathbf{v}_{\rm turb}\cdot\mathbf{v}_{\rm res}dV = \varphi \frac{1}{2}\int_{V_{\rm vir}}\dot{\rho}_{acc}(\mathbf{v}_{\rm res}^2 + \mathbf{v}_{\rm turb}^2)dV = \varphi(\frac{3}{2} \dot{M}_{\rm acc}\sigma_{\rm res}^2 + \frac{3}{2}\dot{M}_{\rm acc}\sigma_{\rm cl}^2 + \gamma \dot{M}_{\rm acc} v_{\rm esc}^2)
\label{varphidef}
\end{equation}
where $\varphi$ is an interpolation parameter that ranges between zero and unity and
\begin{equation}
\gamma = \frac{1}{2}+f\left(\frac{1}{10-4k_{\rho}}-\frac{\chi}{6-2k_{\rho}}\right).
\label{gammadef}
\end{equation}

Summing the individual terms derived above yields the global form of the equation of energy conservation,
\begin{equation}
\begin{split}
\label{energyeqn}
\frac{d\mathcal{E}_{\rm cl}}{dt}&=\frac{\dot{M}_{\rm cl}}{M_{\rm cl}}[\mathcal{E}_{\rm cl}+(1-\eta_{B}^{2})\mathcal{W}]+\frac{GM_{\rm cl}\dot{M}_{\rm cl}}{R_{\rm cl}}\chi\left(1 - \frac{M_{\rm cl}\dot{R}_{\rm cl}}{\dot{M}_{\rm cl}R_{\rm cl}}\right) + \left(\frac{3-k_{\rho}}{4-k_{\rho}}\right)\dot{R}_{\rm cl}(\dot{M}_{\rm ej}v_{\rm ej}^{'}-\xi\dot{M}_{\rm acc}v_{\rm esc})\\
& - 4\pi P_{\rm amb}R_{\rm cl}^{2}\dot{R}_{\rm cl}   - a_{\textrm{I}}\dot{M}_{\rm acc}\dot{R}_{\rm cl}^{2} - 3\dot{M}_{\rm acc}\sigma_{cl}^{2} + \varphi(\frac{3}{2} \dot{M}_{\rm acc}\sigma_{\rm res}^2 + \frac{3}{2}\dot{M}_{\rm acc}\sigma_{\rm cl}^2 + \gamma \dot{M}_{\rm acc} v_{\rm esc}^2)+ \mathcal{G}_{\rm cl}-\mathcal{L}_{\rm cl}
\end{split}
\end{equation}

\noindent
where $\mathcal{L}_{\rm cl} = \int_{V_{\rm vir}} \Lambda dV$.

\end{appendix}

\end{document}